\documentstyle[preprint,aps,prl,psfig]{revtex}
\tighten
\draft
\begin{document}

\draft

\preprint{September 20, 1998}
\title{Computer Simulations of Supercooled Liquids and
Glasses\footnote{To appear as a Topical Review Article in J. Phys.: 
Condens. Matter}}
\author{Walter Kob}
\address{Institut f\"ur Physik, Johannes Gutenberg-Universit\"at,
Staudinger Weg 7, D-55099 Mainz, Germany}
\maketitle

\begin{abstract}
After a brief introduction to the dynamics of supercooled liquids, we
discuss some of the advantages and drawbacks of computer simulations of
such systems.  Subsequently we present the results of computer
simulations in which the dynamics of a fragile glass former, a binary
Lennard-Jones system, is compared to the one of a strong glass former,
SiO$_2$. This comparison gives evidence that the reason for the
different temperature dependence of these two types of glass formers
lies in the transport mechanism for the particles in the vicinity of
$T_c$, the critical temperature of mode-coupling theory. Whereas the
one of the fragile glass former is described very well by the ideal
version of mode-coupling theory, the one for the strong glass former is
dominated by activated processes. In the last part of the article we
review some simulations of glass formers in which the dynamics below
the glass transition temperature was investigated. We show that such
simulations might help to establish a connection between systems with
self generated disorder (e.g. structural glasses) and quenched disorder
(e.g. spin glasses).

\end{abstract}


\noindent
\section{Introduction}
\label{sec0}
The history of man-made glasses is several thousand years
old~\cite{glass_hist} and thus it might seem that we had enough time to
gain an excellent understanding of this type of material. A brief
glance at the recent literature and the topic of many specialized
conferences shows, however, that this is not the case at all. Although
a vast amount of detailed knowledge on various properties of all sorts
of glasses has been accumulated, and our understanding of these
materials has certainly increased tremendously since ancient times, the
answer to some of the most basic issues are still a matter of debate.
Apart from some questions that have been posed already many years ago,
such as regarding the mechanism that gives rise to the glass
transition, new questions have emerged very recently, such as, e.g.,
the nature of the glass transition in small pores, whether or not
glasses are dynamically homogeneous or heterogeneous, or how systems
with frozen in disorder (e.g. spin glasses) are related to the ones
with self generated disorder (e.g. structural glasses). Because of the
vastness of the field the present article does of course not even
attempt to give an exhaustive review on all these different topics and
developments and we refer the interested reader to the various
textbooks and review articles on that
subject~\cite{glass_books,debenedetti97,binder86}. In the following we
will therefore focus on only certain topics of supercooled liquids and
glasses. Hence the fact that many other issues will be treated only
briefly or not at all should not be viewed as a statement of their
irrelevance but rather as a (somewhat arbitrary) choice of the author.

Computer simulations have of course a much shorter history than glasses
since the first work dates back only to the 1950's~\cite{old_sim}.
However, by now it has been shown that such simulations can be an
excellent tool to investigate the properties of complex
systems~\cite{sim_books} and it can be expected that with the
availability of faster and cheaper computers, as well as improved
algorithms, such simulations will play an even more important role in
the future than they do now. In this article we will discuss how such
simulations can be used to gain insight into the structure and the
dynamics of supercooled liquids and glasses. Again, the work on this
subject is by now far too extensive to be completely covered in this
article and therefore we present only a small, but hopefully relevant,
subset of it and refer the interested reader to the mentioned
textbooks~\cite{glass_books,debenedetti97} and other review articles on
this subject~\cite{binder86,glass_sim_rev,sio2_rev}.

In order to facilitate the reading of the articles for those not
familiar with the subject we will give in Sec.~\ref{sec1} a brief
introduction to the field of supercooled liquids and glasses and
discuss some of the questions that are currently debated. The following
section is then devoted to issues related to computer simulations. In
Sec.~\ref{sec3} we will discuss results regarding the {\it equilibrium}
dynamics of fragile and strong glass formers and at the end show that
also in the {\it non-equilibrium} dynamics of glass forming liquids
many very interesting phenomena occur, which so far are not understood
at all, and which can be studied very well by means of computer
simulations.

\section{Supercooled Liquids and Glasses}
\label{sec1}

In this section we will briefly review some facts regarding the
structure and dynamics of supercooled liquids and glasses. Despite its
briefness it should allow the reader to get familiar with some of the
issues concerning supercooled liquids and glasses so that the following
sections become more comprehensible.  More detailed discussions to
these topics can be found in
Refs.~\cite{glass_books,debenedetti97,binder86}.

First of all it is appropriate to specify what we mean in the following
by the terms ``supercooled liquids'' and ``glass''. The standard point
of view is the following: If a liquid can be cooled below its melting
temperature $T_m$ without the occurrence of crystallization, it is
called a good glass former, and when the temperature is less than $T_m$
the system is called supercooled. The static and dynamical properties of
such systems can be studied in a large temperature range below $T_m$
and it is found that their relaxation times increase very fast by many
(12-14) decades if the temperature is lowered. At a certain temperature
the relaxation time exceeds the time scale of the experiment and
therefore the system will fall out of equilibrium. It is this falling
out of equilibrium what is called the glass transition. At temperatures
well below this glass transition temperature no relaxation seems to
take place anymore (on any reasonable time scale) and it is customary
to call this material a glass. (Note that this transition temperature
will in general depend on the type of experiment since its definition
involves the time scale of the experiment. This definition should also
not be confused with the one often used by experimentalists in which a
glass transition temperature is defined as the temperature at which the
viscosity of the system has the (somewhat arbitrary) value $10^{13}$
Poise.)

In the following we will adapt a point of view on ``supercooled'' which
is slightly different and is motivated by the experimental observation
that if a system approaches its glass transition temperature its
relaxation dynamics becomes non-Debye, i.e. that the time correlation
functions decay in a non-exponential way (e.g. they show a two-step
relaxation), whereas it shows a Debye-behavior at high temperatures.
The reason for this phenomenon will be discussed below. It has been
found, however, that, e.g., Glycerol and B$_2$O$_3$ show this sort of
non-Debye behavior also at temperatures significantly {\it above} the
melting temperature~\cite{wuttke94,rossler94}. Therefore one has to
conclude that the non-Debye relaxation has nothing to do with the
system being supercooled, a point of view which is supported from the
theory of the dynamics of dense liquids~\cite{mct}. In addition it is
not hard to imagine a system that does not crystallize at all, i.e. one
for which the concept of a melting temperature, and hence the term
``supercooled'', does not even exist (e.g. in atactic polymers).
Nevertheless it can be expected that the dynamics of such a system will
become very slow when the temperature is decreased and that thus the
system will undergo a glass transition.  For these reasons we will mean
in the following by ``supercooled'' that the relaxation dynamics of the
system is non-exponential and {\it not} that its temperature is below
$T_m$.  Furthermore, in order to simplify the language, we will use in
the following always temperature as the variable that drives the
slowing down of the dynamics. The reader should, however, bear in mind
that there are systems in which the glass transition is driven by a
change of particle concentration, such as colloids or kinetic lattice
gases~\cite{colloids,lattice_gas}.

If the structural and dynamical properties of a good glass former are
measured in the temperature range between the high temperature regime
and the glass transition temperature, one finds that all structural
quantities (density, structure factor, etc.) and thermodynamic
quantities (specific heat, etc.) show a very gentle temperature
dependence, in that they change between a few percent or a factor of
2-3~\cite{glass_books,debenedetti97}.  (Note that the experimental
observation that the specific heat $C_p$ shows a pronounced drop at the
glass transition temperature, see, e.g., Ref.~\cite{bruening92} does
not contradict this statement, since this effect is due to fact that
the system falls out of equilibrium, i.e., that certain translational
degrees of freedom do not contribute anymore to the specific heat. Thus
the drop in $C_p$ is a purely kinetic effect.) As already mentioned
above, dynamic quantities, such as the diffusion coefficient $D$ or the
viscosity $\eta$, will in general show a much more pronounced
temperature dependence than thermodynamic quantities. It is this huge
variation of the transport coefficients that makes the experimental
investigation of the dynamics, and its theoretical description, such an
interesting challenge.  Phenomenologically the temperature dependence
of the dynamics can be described quite well by the so-called
Vogel-Fulcher law $\eta \propto \exp(ET_0/(T-T_0))$, where $T_0<T_g$ is
the so-called Vogel temperature and $E$ is a parameter that determines
the shape of the curve.  If $E$ is large the temperature dependence is
Arrhenius like and if it is small, $\eta$ shows a pronounced upward
bend at a temperature a bit above $T_0$.  Angell coined the terms
``strong'' and ``fragile'' glass formers for the former and latter
case, respectively~\cite{angell85}. Below we will discuss how the
dynamics of strong and fragile glass formers differ on the microscopic
level.

Having discussed some of the phenomena observed in the temperature
dependence of glass forming liquids it might be useful at this point to
make some comments on the various theoretical approaches that have been
used to rationalize the dramatic slowing down of the dynamics when the
temperature is lowered. One of the most simple potential mechanisms is,
that upon cooling the system approaches the critical point of a second
order phase transition and hence the increase of the relaxation times
is just the critical slowing down of the system when the temperature
approaches this point. It is generally believed that it is this
mechanism that is responsible for the glass transition in spin
glasses~\cite{binder86,rieger95,young98,sherrington98}. For structural
glasses the situation is much less clear since the presence of a second
order thermodynamic phase transition implies that one should be able to
identify an order parameter or a growing length scale. Since in
computer simulations it is relatively easy to measure such an order
parameter or length scale, {\it if one knows what to look for}, they
have been used quite extensively to find evidence for the existence of
a thermodynamic transition in structural glasses. However, as discussed
in detail in Ref.~\cite{kob95}, these efforts have led to no positive
results so far.

One of the first successful theories for the glass transition is the one
proposed by Gibbs and DiMarzio~\cite{gibbs_dimarzio58} for dense
polymer melts. The basic idea of this theory is that with decreasing
temperature the configurational entropy to the polymers decreases and
vanishes at a finite temperature, thus leading to the glass transition.
Below we will come back to this theory and discuss it more closely in
the context of some computer simulations that have been performed to
check its validity.

Another very successful theory is the so-called mode-coupling theory
(MCT) which has been worked out by G\"otze, Sj\"ogren and
coworkers~\cite{mct}. Starting from the theory of dense simple
liquids~\cite{theory_liquids} MCT derives equations of motion for
time and wave-vector dependent correlation functions and makes very
detailed predictions on the time and temperature dependence of these
quantities when the system is in the supercooled state. As will be
discussed below, many of the predictions of this theory have been
confirmed in experiments and computer simulations and thus MCT can
presently be regarded as the best available theory of the dynamics of
supercooled liquids.

In the following sections we will extensively discuss computer
simulations that have been done in order to test the validity of the
two theories just mentioned. The goal of this discussion is not to
give an exhaustive review of all the possible tests that have been
done so far, but rather to present some exemplary results of
simulations in order to convince the reader that such simulations can
be a very powerful tool to check to what extend a theory is valid or
not.

\section{Computer Simulations of Glass Formers: Advantages and
drawbacks}
\label{sec2}

The goal of this section is to discuss certain aspects of computer
simulations that are particular to simulations of supercooled liquids
and glasses. More general introductions to simulations can be found
in various textbooks, such as Refs.~\cite{sim_books}.

The most outstanding advantage of computer simulations is that they
provide an extremely large freedom regarding the systems that can be
studied. {\it In principle}, it is no problem to investigate any
Hamiltonian that one is interested in, be it on a classical or quantum
mechanical level. By a judicious choice of the Hamiltonian considered,
it is therefore possible to investigate, e.g., systems in which the
dynamics is determined by purely kinetic constraints instead by
energetic ones~\cite{lattice_gas,ritort,kurchan97}, to study molecules
or polymers with an exactly specified shape and
size~\cite{molecules,lewis_wahnstrom,bennemann98,kammerer}, to
investigate the system in thermodynamics states which are difficult to
realize experimentally, such as negative pressures and high
temperatures~\cite{rustad91,jin93,poole97,badro98,lacks98}, or the
dependence of the structure and the dynamics of a system in a small
pore of a {\it well-defined} size and shape~\cite{baschnagel95,fehr95}.
For example simulations of supercooled water have given evidence that
this system has a liquid to liquid transition~\cite{water_water_trans}
and from simulations of SiO$_2$ novel (crystalline) phases have been
predicted~\cite{badro97}.

In addition to this, simulations offer the unique possibility to access
any observable of interest, since the complete information on the
positions and velocities of all the particles is available at any
given time. This property of simulations allows to determine
quantities which are very difficult to access in real experiments or
very hard to obtain with reasonable precision in an analytical
calculation. Hence it is possible to use such simulations to test
theoretical concepts and theories in a more stringent way than it is
feasible with real experiments. Examples for such type of simulations
will be discussed below.

Despite all these advantages reality is not quite as rosy as it might
seem since one is faced with the sad fact that computer resources are
finite. Therefore it is, e.g., presently out of question to do a full
quantum mechanical simulation for several thousand particles over a time
scale of, e.g., 1~ns, and even with classical force fields and
$O(10^3)$ particles it is hard to simulate times significantly longer
than 100~ns, since each time step is usually only $10^{-15}$ sec long
(for atomic systems). The present state of the art is to make an {\it
ab initio} calculation of about 100 particles for a time span of
10~ps~\cite{sio2_initio,pasquarello98b} whereas for classical
simulations one can deal with box sizes of 50-100\AA and simulate such
a system for 10-100~ns~\cite{large_sim,horbach96}. (Such a simulation
will then usually take the equivalent of several CPU years on a very
good, dedicated workstation.) Of course one might wonder whether it
really is necessary to study systems that are larger than a few hundred
particles, since it is possible to study many aspects of the glass
transition even with such small systems. However, there exist
situations in which large systems are necessary. E.g. the {\it
dynamics} of strong glass formers shows quite pronounced finite size
effects~\cite{horbach96}, and the same is true also for the density of
states in the frequency range which is important for the so-called
boson-peak, a dynamic feature at low energies whose nature is
currently a matter of intense debate~\cite{bp_papers,taraskin97}.

Note that, despite the mentioned limitations in time and system size,
large scale computer simulations cover a wave-vector and frequency
range which is comparable or even larger than, e.g., the one accessible
in neutron-scattering experiments and hence they can be a valuable
addition to investigate the structure and dynamics of systems on this
length and time scale. However, it has to be emphasized that real
experiments still have a crucial advantage over simulations, namely in
the way the sample is prepared. What experimentalist do is to adjust
the temperature of the sample to the temperature of interest and wait
on the order of minutes or even weeks before they start to do the
measurement. Therefore it becomes possible to probe the {\it
equilibrium} dynamics of the system on the ns scale (e.g. in the case
of neutron scattering) even if the relaxation time of the system is on
the order of days. In computer simulations such a procedure is (not
yet) possible. The only way to equilibrate the system is to do a
simulation at the temperature of interest for a time span that is on
the order of the longest relevant relaxation time of the system and to
subsequently start the run for the production. Since typical time spans
of the simulation do not exceed 100~ns, it is therefore only possible
to equilibrate the system at a relatively high temperature. Thus all
equilibrium measurements are restricted to these temperatures as well.

In order to circumvent this problem one might be tempted to quench the
system relative rapidly to a temperature at which its relaxation time
is larger than the one accessible to computer simulations, let the
system relax for some time, and subsequently start to measure its
properties. Such an approach is, however, quite dangerous in that the
results from such a simulations will in general show aging effects.
This means that quantities that should be constant (such as the average
potential energy) show a small drift, and that time correlation
functions will no longer be time-translation invariant. Some of
these effects will be discussed in more detail in Sec.~\ref{sec3b}.

We stress, however, that the above mentioned problem with the
equilibration of the sample is not a principle one. There is no reason
why a cleverly designed Monte Carlo algorithm should not be able to
equilibrate the sample also at a temperature at which, e.g., the
relaxation time {\it for the usual Newtonian dynamics} is
macroscopically large.  Examples for such algorithms have already been
successfully implemented for polymers and it was found that they allow
to equilibrate the sample about 100 times faster than with the normal
dynamics~\cite{sokal95}. In recent years other methods have been
proposed and tested and, although presently there still is no optimal
method in sight, the progress is quite encouraging~\cite{new_algorithms}
and therefore it can be hoped that sooner or later this bottleneck will
be removed.

Before we end this section it is appropriate to make some comments on
the Hamiltonians used in simulations of supercooled liquids and
glasses. Roughly speaking simulations in this field can be divided into
two types: 

i) In the first one the main goal is to use the simulation to gain an
understanding on some very general question, such as to what extend MCT
gives a correct description of the dynamics, the nature of the
dynamical heterogeneities observed in
experiments~\cite{dyn_het_exp,dyn_het_sim}, or the search for a
diverging length scale when the temperature approaches
$T_g$~\cite{baschnagel95,length_scale}.  Since one is interested in
some universal properties of glassy systems, such simulations are
usually done with very simple systems, such as Lennard-Jones particles,
kinetic Ising models~\cite{lattice_gas} or even simpler ones (see,
e.g., the backgammon model of Ritort~\cite{ritort}). Because of their
simplicity such models allow to obtain results of a much higher
accuracy than it would be possible with more realistic and thus more
complicated systems and thus more definite answers can be given to the
question of interest.

ii) In the second type of simulation one attempts to answer some quite
specific questions for a given material (or class of materials), such
as, e.g., the nature of the so-called boson-peak in strong glass
formers~\cite{bp_papers,taraskin97,habasaki95,horbach98b,delanna98}, to
identify the mechanism leading to the so-called ``mixed alkali
effect''~\cite{mixed_al_effect}, or to determine the distribution
function of the rings in network forming systems such as
SiO$_2$~\cite{pasquarello98b,rino93,vollmayr96a,vollmayr96b}. For such
simulations is is important to have a reliable potential at hand which
is able to give a sufficiently realistic description of the quantity
that one is interested in.  Unfortunately it is still the exception
rather than the rule that for the material of interest a good potential
is available.  The reason for this is that the development of a
reliable force field often involves a substantial amount of {\it ab
initio} calculations, in order to determine the potential energy of
some typical configurations, and also the subsequent fit of these data
points to a classical potential energy function is often rather
difficult~(see, e.g., the papers of Takada {\it et
al.}~\cite{takada95}). Both of these steps require a substantial amount
of expertise and work and thus have been done only for a few selected
substances, such as, e.g.,
silicates~\cite{tsuneyuki88,beest90,kramer91,gonze94} or
ZnCl$_2$~\cite{wilson94}.

\begin{figure}[f]
\psfig{file=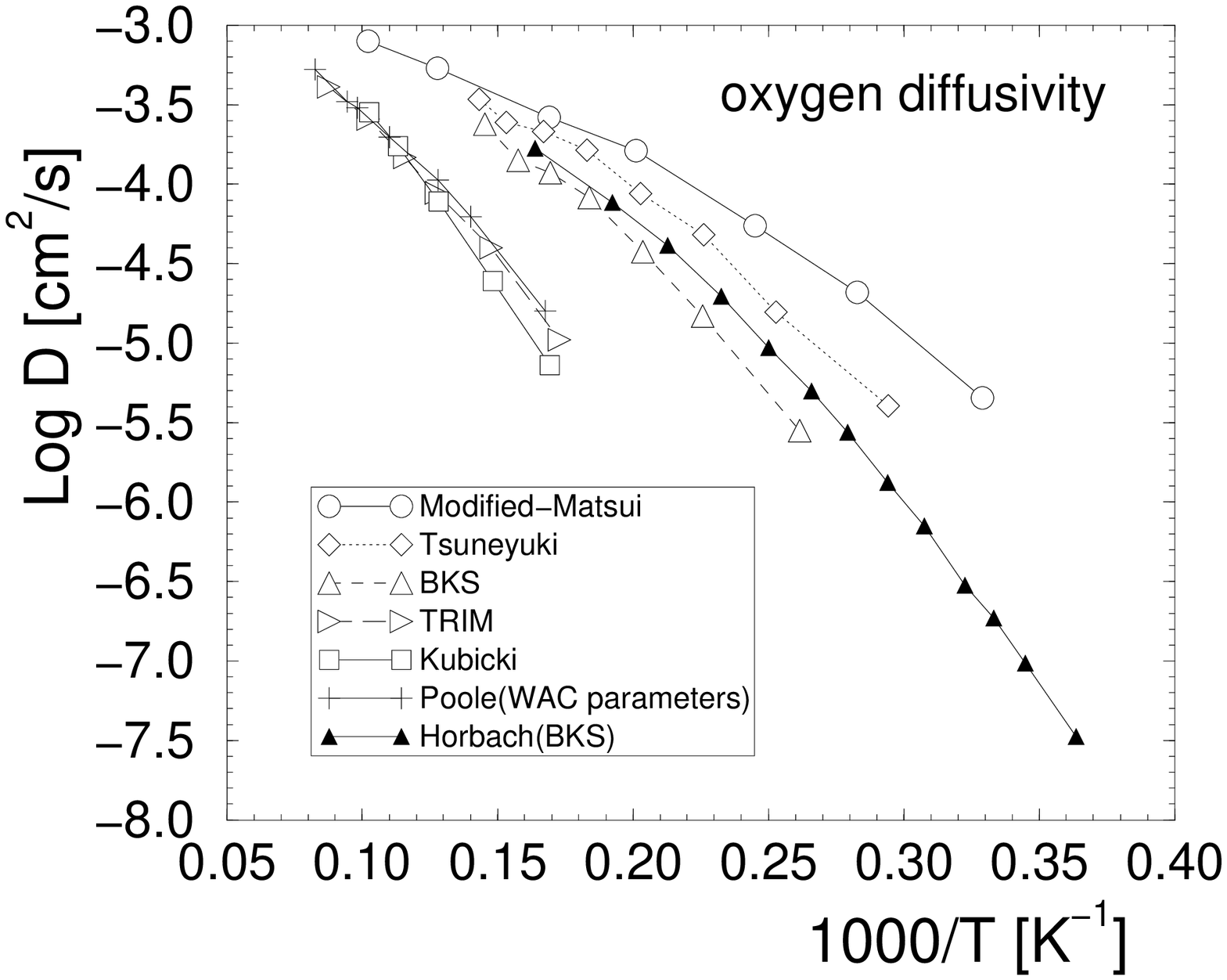,width=13cm,height=9.5cm}
\caption{Temperature dependence of the oxygen diffusion constant for
different models of SiO$_2$. See Ref.~\protect\cite{hemmati98} for
details. Adapted from Ref.~\protect\cite{hemmati98}, with permission.}
\label{fig1}
\end{figure}

It should also be noted that whether or not a potential is realistic
depends on the question one is interested in. For the case of silica,
e.g., there exist many different potentials, many of which give a quite
realistic representation of the {\it structural} properties of
amorphous SiO$_2$, such as the structure
factor~\cite{sio2_rev,rustad91,badro97,taraskin97,rino93,vollmayr96a,vollmayr96b,hemmati98,vessal89,sio2_sim,horbach98a}.
Thus from this point of view the various potentials can be considered
as essentially equivalent.  If, however, {\it dynamical} quantities are
considered, such as the diffusion constant, one finds that the various
potentials make very different predictions. In Fig.~\ref{fig1} we
reproduce data by Hemmati and Angell in which the temperature
dependence of the oxygen-diffusion constant is shown for different
potentials~\cite{hemmati98}. One sees that at high temperatures the
values of the diffusion constants predicted by the different potentials
differ only by about a factor of three.  This agreement changes
dramatically when the temperature is lowered in that the different
models predict diffusivities that vary over several orders of
magnitude. Hence we conclude that a potential that might be appropriate
to describe the structure of SiO$_2$ might be unsuitable for describing
the dynamics of the system.

\section{Some Examples of Computer Simulations of Supercooled Liquids 
and Glasses}
\label{sec3}

In this section we will discuss some computer simulations of
supercooled liquids and glasses. In the first part we will make a
comparison between the {\it equilibrium} dynamics of a fragile glass
former and a strong glass former in order to see what the similarities
and differences in these two types of systems are. In the second part
we will discuss the {\it non-equilibrium} (aging) dynamics of a simple
glass former, i.e.  the decay of the time correlation functions of the
system after it has been subjected to a quench to a low temperature.

\subsection{The relaxation dynamics of strong and fragile glass
formers}
\label{sec3a}

We already mentioned at the beginning of this article that the temperature
dependence of transport quantities of glass forming liquids is not
universal in that certain glass formers (called ``strong'') show an
Arrhenius dependence and others (called ``fragile'') show strong
deviations from it~\cite{angell85}. In this subsection we will present
the results of computer simulations in which the dynamics of a simple
fragile glass former and SiO$_2$, a prototype of a strong glass former
was investigated. By a comparison between the two dynamics we will
attempt to understand what the difference between the two types of
glass formers are {\it on the microscopic level}.

Very often fragile glass formers are van der Waals liquids in which the
interactions between the (small) molecules is relatively weak.  Thus
such systems can be modelled by particles with isotropic short range
interactions such as soft spheres, $V(r) \propto r^{-12}$, or the
Lennard-Jones potential. Because of the simplicity of these potentials
the relaxation dynamics of such models has been studied extensively by
means of computer
simulations~\cite{kob95,ullo,hansen,hiwatari,hurley,parisi_sim,parisi97,wahnstrom,kob_lj,kob95b}.
Early simulations often focussed on one-component systems~\cite{cape80}
but with the increased speed of the computers it was found that such
systems started to crystallize within the time span of the simulation.
A simple way to avoid this problem is to use a binary mixture of
particles and if the interaction parameters and the concentration of
the species is chosen well, such systems stay in the (meta)stable
liquid-like state for all time spans which a present state of the art
simulation can cover, i.e.  over 100 million time steps, which
corresponds to about 100~ns. A binary mixture of soft spheres has,
e.g., been investigated in great detail by the group of
Hansen~\cite{hansen}, work that has been continued by Hiwatari, Odagaki
and coworkers~\cite{hiwatari}, and very recently this system has been
used by Parisi and coworkers to investigate aging
phenomena~\cite{parisi_sim,parisi97}.

The binary system we will discuss here consists of an 80:20 mixture of
Lennard-Jones particles, $V_{\alpha\beta}(r)=4\epsilon_{\alpha\beta}
[(\sigma_{\alpha\beta}/r)^{12}-(\sigma_{\alpha\beta}/r)^6]$, where
$\alpha, \beta \in \{A,B\}$ denotes the type of the particle ($A$ being
the majority species).  The parameters of the potential are given by
$\sigma_{AA}=1.0$, $\epsilon_{AB}=1.5$, $\sigma_{AB}=0.8$,
$\epsilon_{BB}=0.5$, and $\sigma_{BB}=0.88$. For this system we will
report length in units of $\sigma_{AA}$, and energy and time in units
of $\epsilon_{AA}$ (setting $k_B=1$) and
$(m\sigma_{AA}^2/48\epsilon_{AA})^{1/2}$, respectively, where $m$ is
the mass of the particles. In the simulation we used a cubic box, of
length 9.4, with periodic boundary conditions and the total number of
particles was 1000, which is large enough to avoid finite size effects
almost completely. More details on these simulations can be found in
Refs.~\cite{kob_lj,kob95b}.

As already mentioned above, there exist quite a few different
potentials for SiO$_2$~(see, e.g., the references
in~\cite{vollmayr96b}). The one proposed by van Beest,
Kramer and van Santen (BKS)~\cite{beest90} seems to be one of the
best, in that it is able to reproduce well many of the structural and
dynamical features of real silica. In this potential the interactions
$\phi_{ij}$ between ions $i$ and $j$ at a distance $r$ apart is given by:
\begin{equation}
\phi_{ij}(r)=\frac{q_i q_j e^2}{r}+A_{ij}\exp(-B_{ij}r)-
\frac{C_{ij}}{r^6}\quad .
\label{eq1}
\end{equation}
Here $e$ is the charge of an electron and the constants $A_{ij}$,
$B_{ij}$ and $C_{ij}$ can be found in Refs.~\cite{vollmayr96b,beest90}.
One interesting aspect of this potential is that it contains only
two-body terms. This is somewhat surprising since SiO$_2$ forms an
open  tetrahedral network and thus it might be expected that three-body
terms are needed as well. However, it has been shown that the BKS
potential is indeed able to generate such a (disordered) tetrahedral
network~\cite{vollmayr96b}, since the competition between the different
two-body forces mimic the three-body forces. As a side remark we
mention that the absence of the three-body terms is of course
advantageous for computer simulations, since their evaluation is
usually demanding from a computational point of view. Thus the only
part of the potential whose calculation is computationally intensive is
the long-range Coulombic part, which is usually calculated by means of
the Ewald summation~\cite{sim_books,deserno98}. Since the computational
effort of this method scales with $N^{3/2}$, where $N$ is the number of
particles, it is clear that doing simulations of large systems with
long range potentials is computationally much more demanding than if
the potential is short ranged (and whose computational effort thus
scales like $N$). However, since a few years new methods have become
available in which the calculation of the long ranged forces also
(essentially) scales like $N$~\cite{fast_coulomb}.  The disadvantage of
these methods is that they become efficient only at relatively large
$N$, and thus for small systems (a few hundred particles) the Ewald
summation is presently the only real possibility to handle such forces.
In practice this means that for a system of the order of $10^3$
particles the calculation of the forces on all the particles in a
system with Coulombic interactions is about one order of magnitude more
time consuming than in the case of short range interactions (e.g.
Lennard-Jones).

Since it has been found that the dynamics of strong glass formers shows
quite large finite size effects~\cite{horbach96} it is necessary to use
for such simulations relatively large systems sizes. In the following
we will discuss results of simulations in which 8016 ions have been
used, which corresponds to a box size of about 48\AA. More details on
the simulation can be found in
Refs.~\cite{horbach96,horbach98a,horbach_diss}.

\begin{figure}[h]
\psfig{file=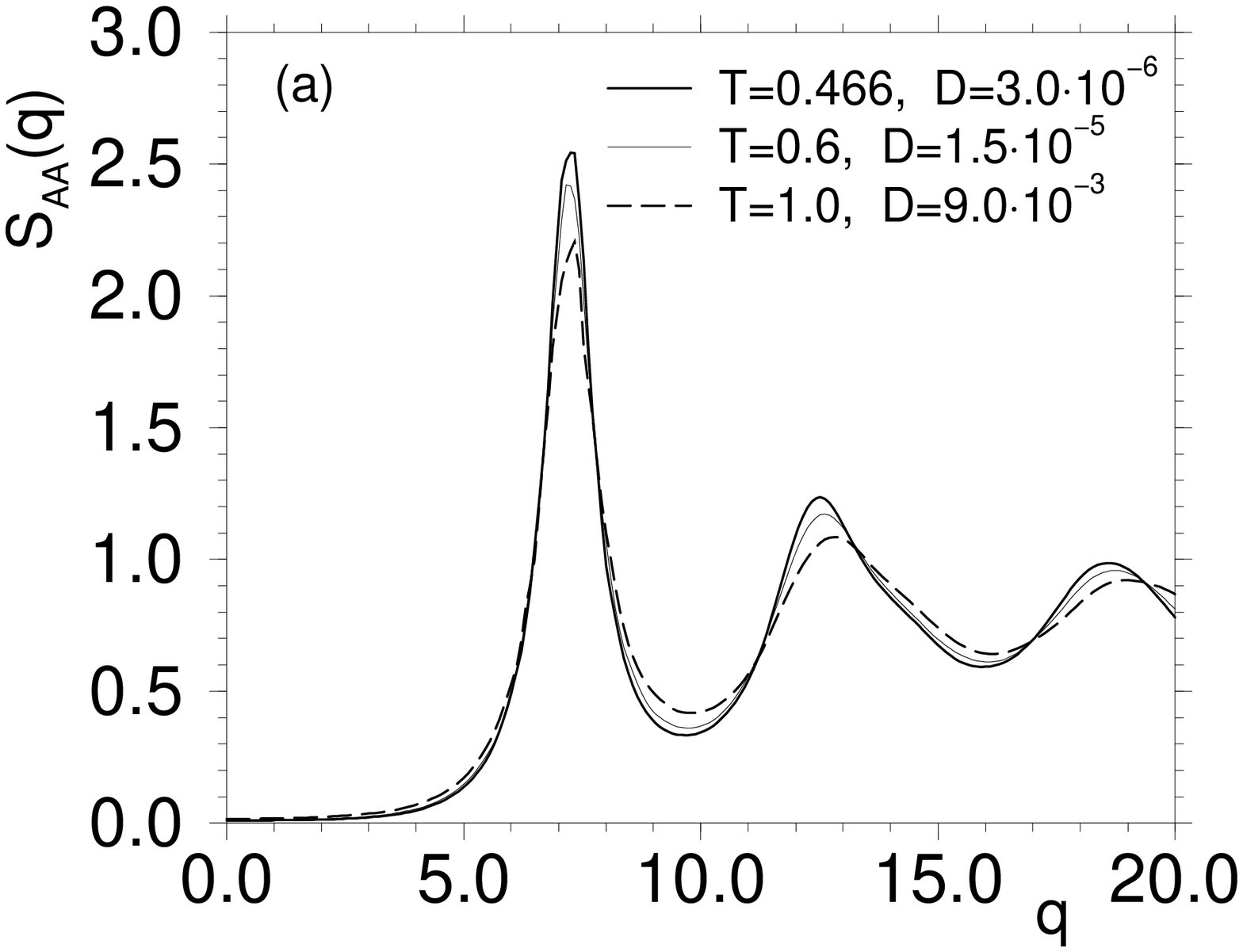,width=13cm,height=9.5cm}
\psfig{file=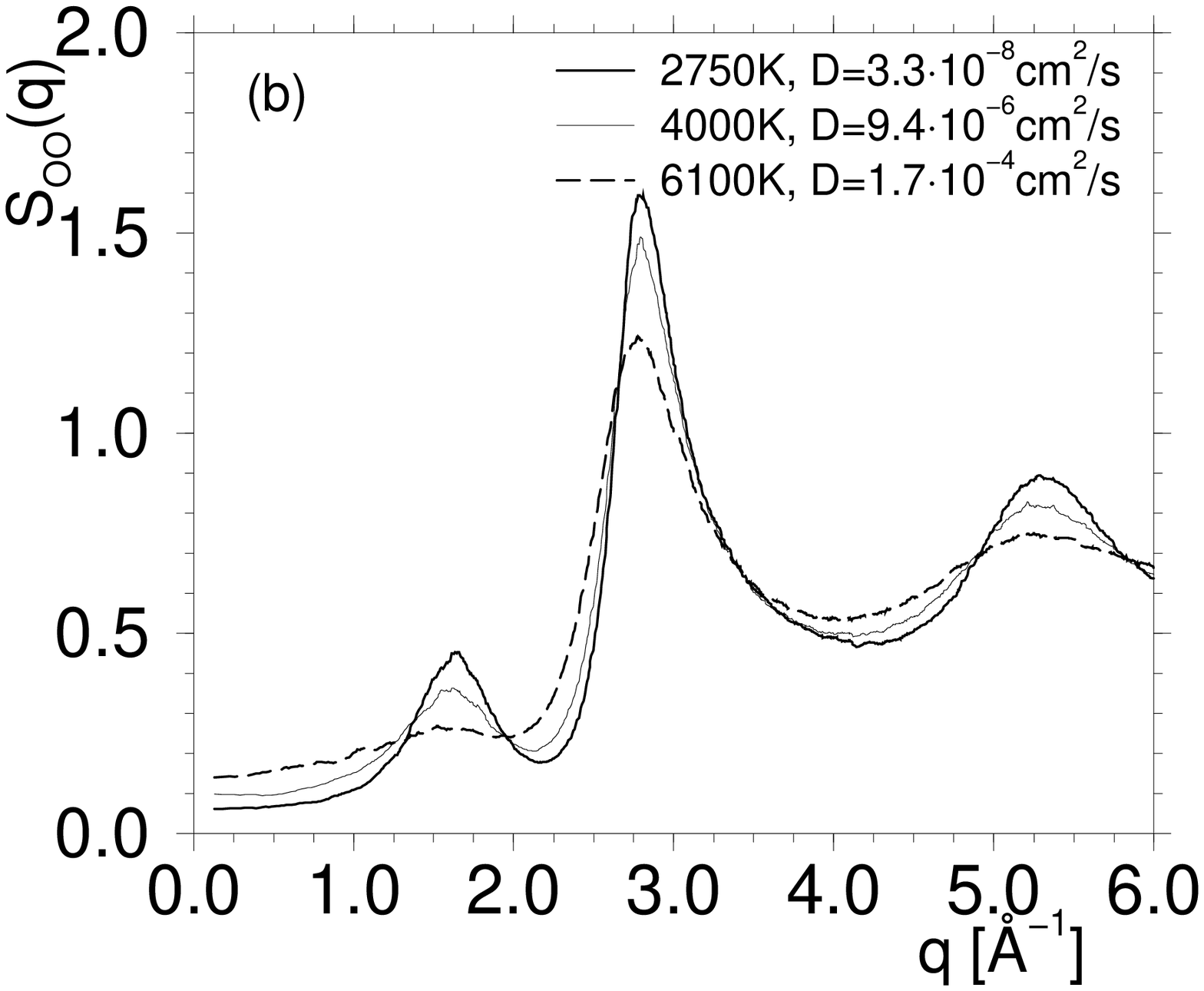,width=13cm,height=9.5cm}
\caption{Wave-vector dependence of the partial static structure factor
for (a) a Lennard-Jones system, a fragile glass former, ($A$-$A$
correlation) and (b) SiO$_2$, a strong glass former
(oxygen-oxygen-correlation). The curves correspond to different
temperatures and range from the temperatures at which the system is in
a normal liquid state, to temperatures at which it is in a deeply
supercooled liquid state. Also given are the values of the diffusion
constants for the $A$ particles (a) and the oxygen atoms (b). 
From Refs.~\protect\cite{kob_lj,horbach98c}.}
\label{fig2}
\end{figure}

Before we start the discussion of the {\it dynamical} behavior of the
two glass formers, we will have a brief look at the temperature
dependence of the {\it static} structure of these systems. One
possibility to do this is to investigate the partial structure factors
$S_{\alpha\beta}(q)$ which are given by $\langle \rho_\alpha(q)
\rho_\beta^*(q) \rangle$, where $\rho_{\alpha}(q)$ is the fluctuation
of the density of particles of type $\alpha$ for wave-vector $q$, i.e.
$\rho_{\alpha}(q)=\sum_{j=1}^{N_{\alpha}} \exp(i {\bf q} \cdot {\bf
r}_j^{\alpha})$, and ${\bf r}_j^{\alpha}$ is the position of the $j$th
particle of type $\alpha$.  In Fig.~\ref{fig2} we show the partial
structure factors for the $A-A$ and oxygen-oxygen correlations, for the
Lennard-Jones and silica system, respectively.  The temperatures for
the different curves range from temperatures at which the system is in
its normal liquid state to temperatures at which the system is in a
deeply supercooled state. From the figure it can be seen that the
temperature dependence of this structural quantity is rather weak and
that the main effect is that the various peaks and minima become more
pronounced when the temperature is lowered. Although the static
structure factors do not depend strongly on temperature, time dependent
quantities do show a strong dependence. In order to demonstrate this we
have included in the figure also the diffusion constant at the
different temperatures. From these figures one recognizes that in the
temperature range considered, the dynamics of the system does indeed
slow down quite dramatically.  We will discuss this point in more
detail below.

One comment on the meaning of the various peaks might be useful. In the
case of a simple liquid, Fig.~\ref{fig2}a, the first peak in the
structure factor corresponds to the typical inter-particle distance.
This is in contrast to the case of network-forming systems,
Fig.~\ref{fig2}b, where this distance corresponds to the {\it second}
peak in $S(q)$. The first peak in the structure factor, often called
first sharp diffraction peak~\cite{glass_books}, is related to the size
of the structural units making up the network (i.e.  tetrahedra in the
case of SiO$_2$).  It is interesting to see that this peak starts to be
visible already at the highest temperature investigated. This means
that the network is forming already at temperatures much higher than
the melting temperature, which is around 2000~K. This is probably the
reason why network-forming systems are often very viscous even at very
high temperatures.

It has to be emphasized that the observed (unspectacular) temperature
dependence of the structure factor is not exceptional for {\it
structural} quantities. Although several efforts have been made to find
a structural quantity that shows a more pronounced temperature
dependence, no clear evidence has been found so
far~\cite{baschnagel95,length_scale}.  Thus the point of view that the
slowing down of the dynamics of {\it structural} glasses is related to
a second order phase transition, and hence to the existence of a
divergent length scale, is so far not supported by good evidence.
(See also Ref.~\cite{weber97} on this point.)

\begin{figure}[h]
\psfig{file=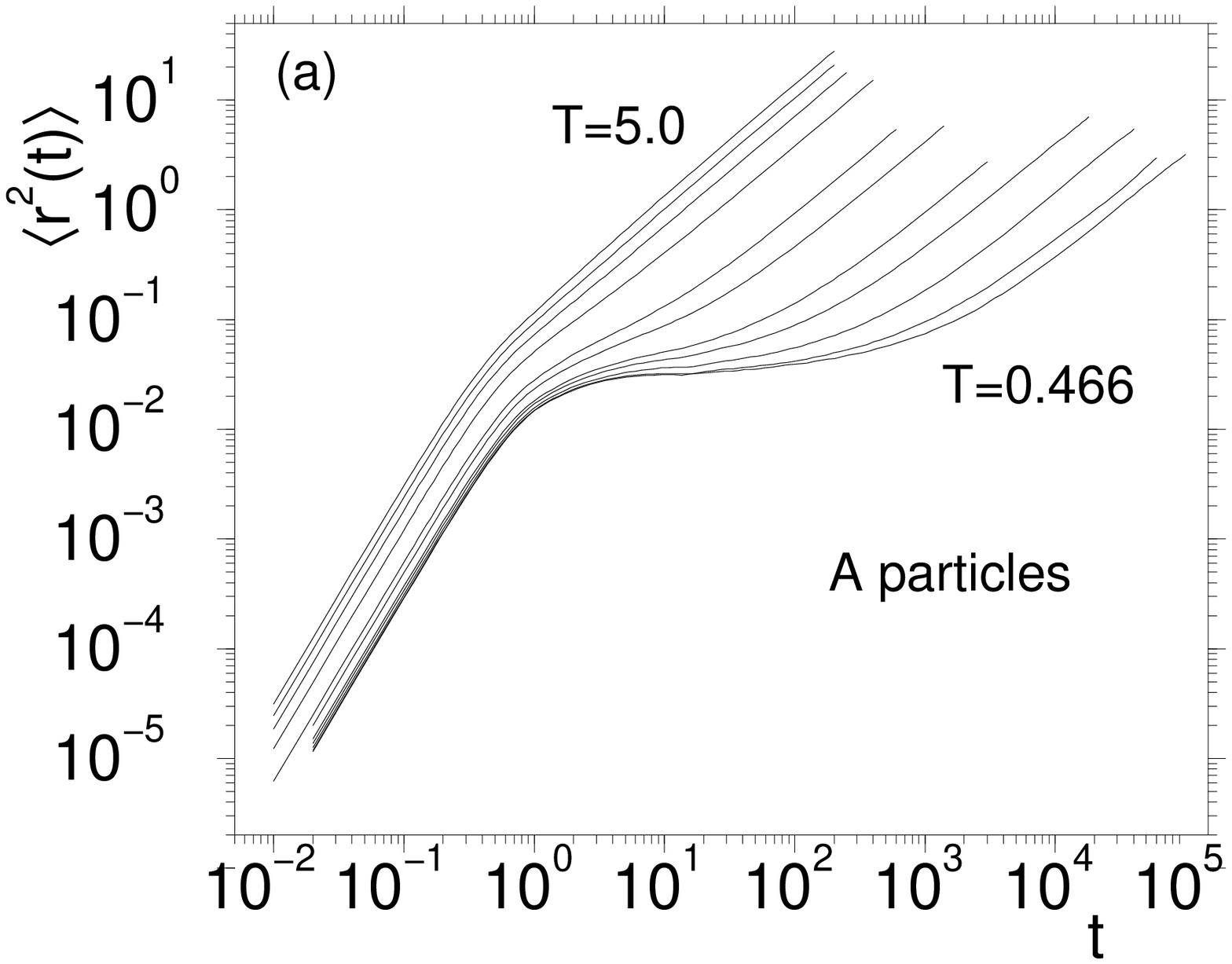,width=13cm,height=9.5cm}
\psfig{file=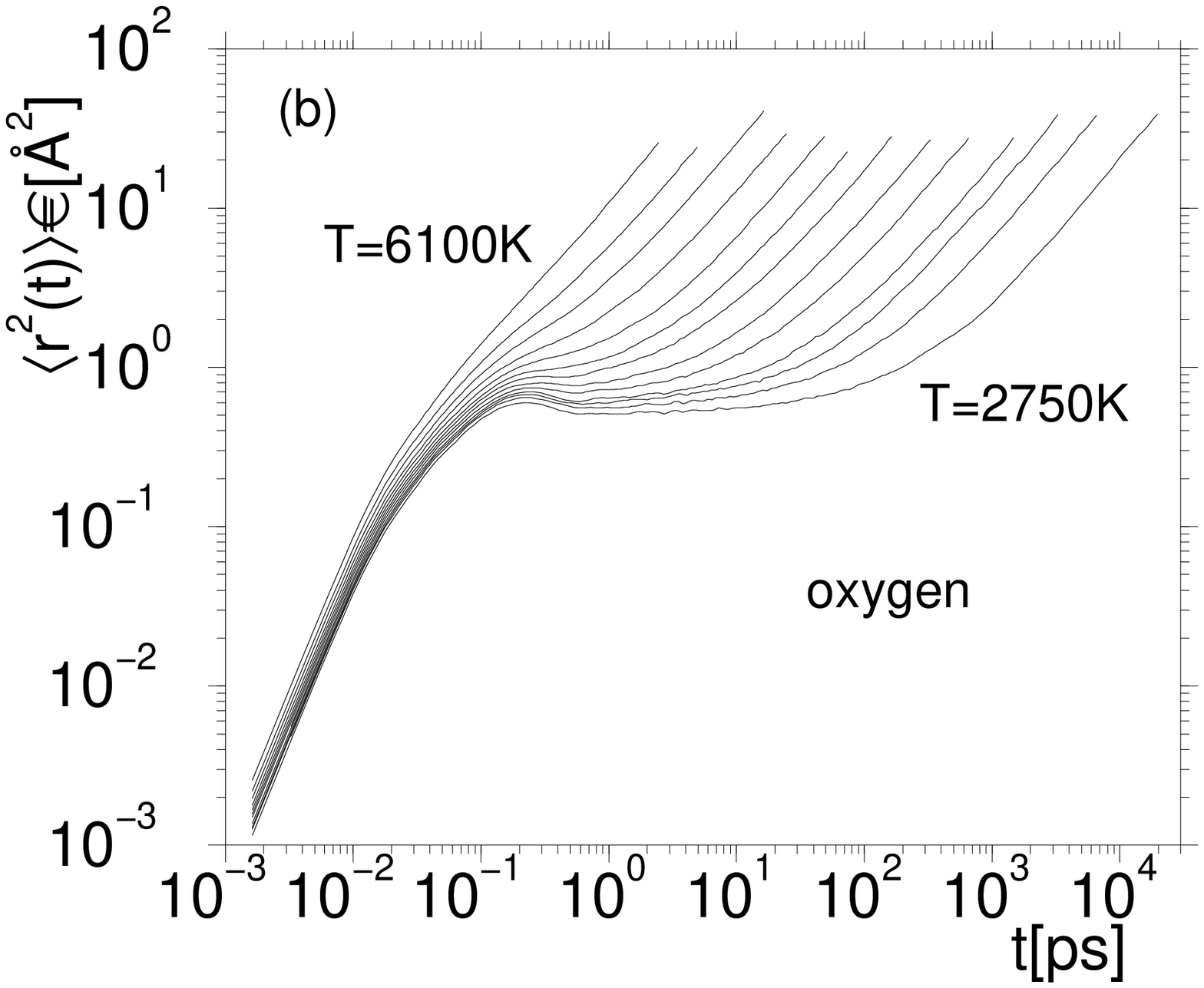,width=13cm,height=9.5cm}
\caption{Time dependence of the mean squared displacement for different
temperatures. a) Lennard-Jones system. $T=5.0$, 4.0, 3.0, 2.0, 1.0,
0.8, 0.6, 0.55, 0.5, 0.475, 0.466. b) Silica. $T=6100$~K, 5200~K,
4700~K, 4300~K, 4000~K, 3760~K, 3580~K, 3400~K,
3250~K, 3100~K, 3000~K and 2900~K and 2750~K. From
Refs.~\protect\cite{horbach98a,kob95b}.}
\label{fig3}
\end{figure}

A much more interesting dependence on temperature than the one for
structural quantities is found in time dependent correlation
functions, or transport coefficients. The simplest example is $\langle
r_{\alpha}^2(t) \rangle$, the mean squared displacement (MSD) of a
tagged particle of type $\alpha$:

\begin{equation}
\langle r_{\alpha}^2(t)\rangle =N_{\alpha}^{-1} 
\sum_{i=1}^{N_{\alpha}} \langle
\delta(r^2-|{\bf r}_i^{\alpha}(t)-{\bf r}_i^{\alpha}(0)|^2)\rangle \quad,
\label{eq2}
\end{equation}
where $\langle . \rangle$ is the thermal average. The time dependence
of the MSD for different temperatures is shown in Fig.~\ref{fig3}. Let
us first consider the MSD for the Lennard-Jones system. At high
temperatures, top curves, the MSD shows at short times a quadratic
dependence on time, $\langle r_{\alpha}^2(t) \rangle \propto t^2$. This
behavior can be understood immediately by realizing that for short
times the particles will move ballistically, i.e. ${\bf
r}_i^{\alpha}(t) \approx {\bf r}_i^{\alpha}(0) +{\dot{{\bf
r}}}_i^{\alpha} t$, and thus give rise to the observed time dependence
for $\langle r_{\alpha}^2(t) \rangle$. For longer times the particles
start to collide with their neighbors and their motion becomes
diffusive. Therefore the MSD shows a linear dependence on time, as can
be seen in Fig.~\ref{fig3}. For low temperatures (bottom curves) the
situation at short and very long times is similar to the one at high
temperatures in that the ballistic and the diffusive behavior are
observed. For intermediate times, however, the MSD shows a feature not
present at high temperatures, namely a plateau.  This means that there
exists a time range, which at the lowest temperatures extends over
several decades, in which the MSD does not increase substantially. The
microscopic reason for this plateau is that the tagged particle is
trapped in the cage formed by the neighboring particles that surround
it, and it takes the particle a long time to escape this cage. Note
that the particles forming this cage are of course sitting in cages as
well and thus the motion of all particles is slowed down.  With
decreasing temperature the cages become more and more rigid and thus
the time needed to break them up increases.  The earlier mentioned
MCT~\cite{mct} is an attempt to describe this breaking up {\it in a
self consistent way} and hence to rationalize the dramatic increase of
the relaxation time on a microscopic level.

In Fig.~\ref{fig3}b we show the MSD for the oxygen atoms in the silica
melt. We recognize that for this strong glass former the curves 
look qualitatively similar to the ones of the fragile glass former. The
main difference is that at low temperatures the MSD for silica shows a
little bump at around 0.2ps~\cite{horbach98a,angell94}. The
reason for this feature lies in the so-called boson-peak, a dynamical
feature whose intensity seems to be related to the fragility of the
glass~\cite{rossler94}, and which will be discussed in more detail
below.

Using the Einstein relation $D=\lim_{t\to \infty} \langle r^2(t)
\rangle/6t$, the diffusion constant $D$ can be calculated from the
MSD.  Before we discuss the temperature dependence of $D$ it is useful
to review a few of the predictions of MCT~\cite{mct} since they will be
helpful to understand the following results. As already mentioned
before, this theory attempts to describe the dynamics of supercooled
liquids in a self-consistent way.  Most of the predictions of the
theory have been worked out for that version of MCT in which certain
terms in the equations of the theory, the so-called hopping terms, are
neglected. This special case is called ``ideal MCT'' and it is
predicted that there exist a special ``critical'' temperature $T_c>0$
in the vicinity of which the dynamics shows an anomalous temperature
dependence, in that the relaxation times $\tau$, or the inverse of the
diffusion constants, show a power-law divergence, $\tau \propto D^{-1}
\propto (T-T_c)^{-\gamma}$.  In this ideal case the system does not
relax anymore if the temperature is below $T_c$. If the mentioned
hopping terms are taken into account the divergence does not really
take place. If these terms are weak the relaxation times will show the
mentioned power-law, but slightly above $T_c$ they will change over to
an Arrhenius law~\cite{hopping}. Empirically it is found that fragile
(and not so fragile) glass formers can be described well by the
idealized MCT~\cite{mct}, whereas not too much is known about strong
glass formers.

\begin{figure}[h]
\psfig{file=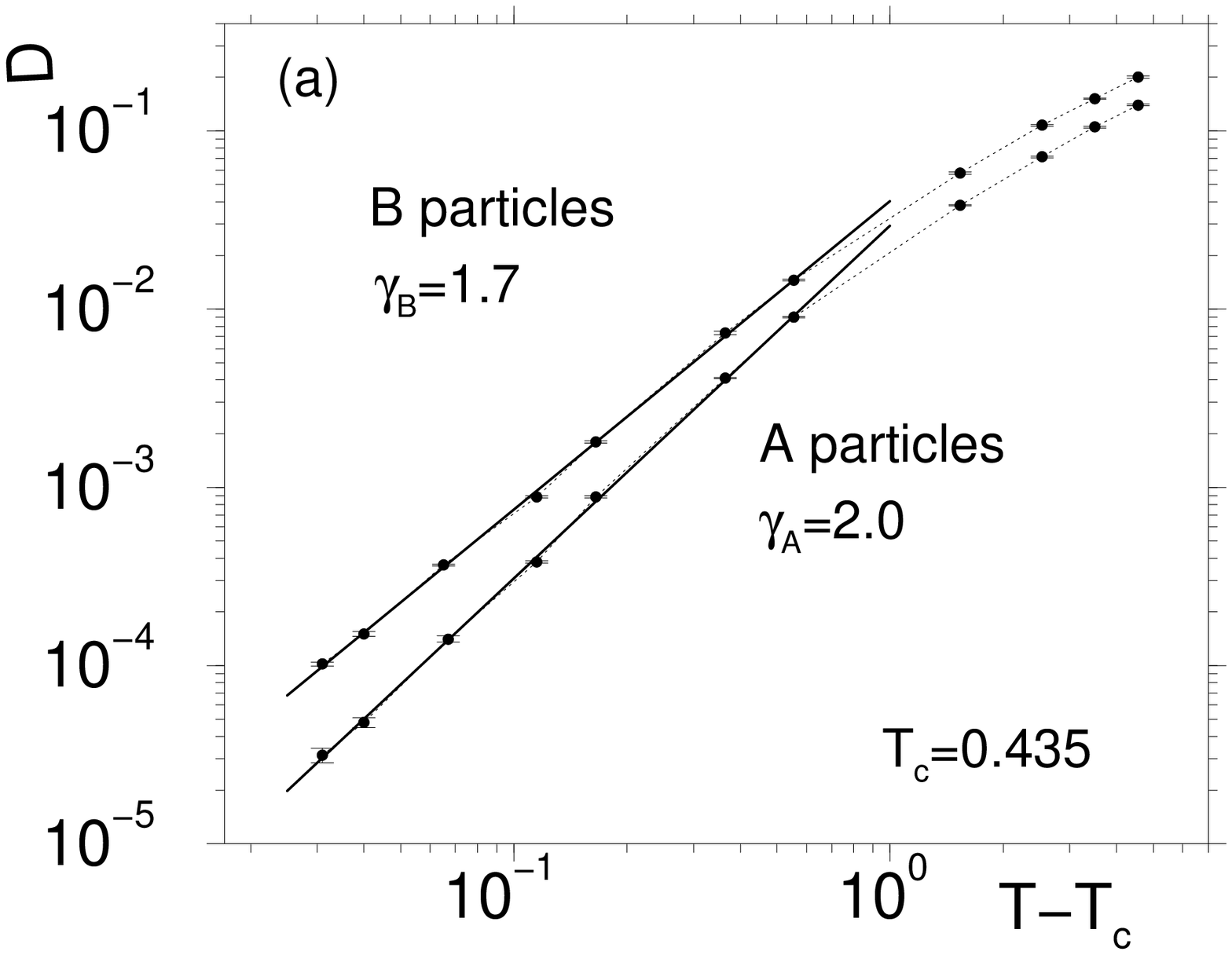,width=13cm,height=9.5cm}
\psfig{file=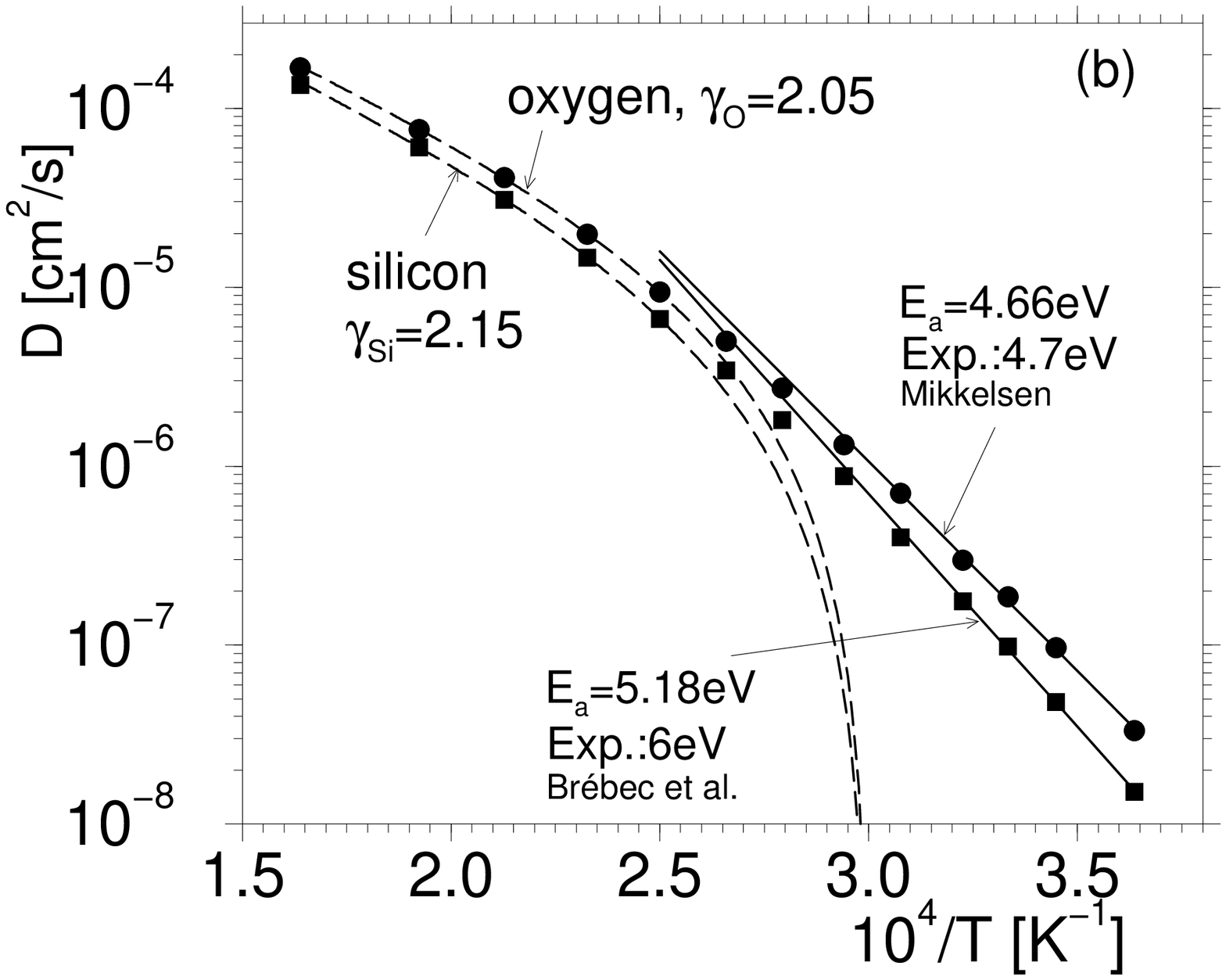,width=13cm,height=9.5cm}
\caption{Temperature dependence of the diffusion constants. (a)
Lennard-Jones system. The solid lines are fits with a power-law of
the form $D\propto (T-T_c)^{\gamma}$. (b) Silica. The solid lines are
fits with an Arrhenius law with the stated activation energies. The
experimental values are from
Refs.~\protect\cite{mikkelsen84_brebec80}. The 
dashed lines are power-law with $T_c$=3330K.
From Refs.~\protect\cite{horbach98a,kob_lj}.}
\label{fig4}
\end{figure}

We now discuss the temperature dependence of the diffusion constants,
which are shown in Fig.~\ref{fig4}. Note that we use two different
types of plots to present the data for the Lennard-Jones and the silica
system. For the former system, a fragile glass former, we have fitted
the low-temperature data with a power-law in order to check whether the
mentioned temperature dependence predicted by MCT gives a good fit to the
diffusion constant. Using $T_c$ as a fit parameter we find that this is
indeed the case (see Fig.~\ref{fig4}a). The critical temperature $T_c$ is
the same for the $A$ and $B$ particles, in agreement with the prediction
of MCT. According to the theory, also the value of the exponent $\gamma$
should be independent of the particle species, and we find that this is
reasonably well fulfilled in that for our system the two value agree to
within 15\% (see figure). 

We also mention that such power-laws have also been found in other
simulations of simple liquids~\cite{lewis_wahnstrom,hansen,wahnstrom},
polymeric systems~\cite{bennemann98}, a simple molecular
liquid\cite{kammerer}, and water~\cite{sciortino}. The result on water
is of particular interest since H$_2$O is a network forming liquid,
thus structurally very similar to silica, the system we will discuss
next.

For silica the temperature dependence of $D$ is more complicated than
the one of fragile glass formers. Since this is a strong glass former,
we expect that at low temperatures an Arrhenius behavior is found and
thus it is reasonable to plot the data versus $1/T$, which is done in
Fig.~\ref{fig4}b. From the figure we see that at low temperatures the
expected Arrhenius law is indeed observed and that the activation
energies are close to the ones found in
experiments~\cite{mikkelsen84_brebec80} (see figure).  Thus we have
evidence that our model for silica is quite realistic not only
regarding static quantities~\cite{vollmayr96b,horbach_diss}  but also
dynamical ones, at least at low temperatures.

At higher temperatures strong deviations from the Arrhenius law are
observed in that the temperature dependence is weaker than the one
expected from the activated dynamics at low $T$. In this temperature
range the curves can be fitted well with a power-law, as predicted by
MCT, with a critical temperature which is independent of the species
and has a value of about 3330~K (see figure). (We also mention that the
given values of $\gamma$ are compatible with the one from the MCT
analysis of the $\beta$--relaxation regime.) Thus from the temperature
dependence of the diffusion constants we have evidence that also the
{\it strong} glass former SiO$_2$ shows at high enough temperatures the
non-Arrhenius dependence observed for fragile glass formers. This
result is in agreement with findings of R\"ossler and
Sokolov~\cite{rossler96}.  By analyzing experimental viscosity data,
these authors arrived to the conclusion that for all glass formers
there exist a temperature range in which the temperature dependence of
the transport coefficients is non-Arrhenius.  Hence we conclude that
the main difference between strong and fragile glass formers is that in
the former the hopping processes that destroy the dynamical singularity
predicted by the ideal version of MCT are so strong, that the signature
of this singularity, namely the power-laws, accounts only for a
relatively small range in the diffusion constant (or other transport
quantities) before the hopping processes take over and lead to an
activated dynamics. In contrast to this are the fragile systems in
which the mentioned power-law can be observed over several decades in
the diffusion constant.

This point of view is also corroborated by the analysis of the self
part of the van Hove correlation function
$G_s(r,t)$~\cite{theory_liquids}. This function, or rather $4\pi r^2
G_s(r,t)$, gives the probability that a particle has moved a distance
$r$ in time $t$. It is found that for fragile glass formers, such as
the present Lennard-Jones system, this function decreases monotonically
as a function of $r$~\cite{kob95b} whereas for silica at intermediate
and low temperatures the distribution function for the oxygen atoms
show a secondary peak at a value of $r$ which corresponds to the
typical oxygen-oxygen distance (no secondary peak is found for the silicon
atoms)~\cite{horbach_diss}. From the existence of such a secondary peak
it has been concluded~\cite{hansen,wahnstrom} that the transport
mechanism is not the flow-like motion described by MCT but more a
jump-like motion that is activated. This point of view is thus in
agreement with the one put forward above for the motion of the atoms in
silica.

Experimentally it is not possible to measure, {\it for atomic systems},
time and space correlation functions like $G_s(r,t)$. However, in
neutron time of flight measurements it is possible to study its space
Fourier transform, the incoherent intermediate scattering function
$F_s(q,t)$~\cite{theory_liquids}, 

\begin{equation}
F_s(q,t)=\frac{1}{N_{\alpha}} \sum_{j=1}^{N_{\alpha}}\langle 
\exp(i {\bf q} \cdot [{\bf r}_j^{\alpha}(t)- {\bf r}_j^{\alpha}(0)] \rangle,
\label{eq3}
\end{equation}
and in neutron and light scattering experiments one has access to its
space and time Fourier transform, the dynamical structure factor
$S(q,\omega)$.  Therefore it is interesting to calculate these
quantities from the simulations as well. (We note, however, that an
{\it accurate} calculation of $S(q,\omega)$ from a simulation is rather
difficult, since calculating the time Fourier transform of a
correlation function that extends over 6-8 decades in time is not a
simple task. Therefore results are usually presented in the time
domain.)

\begin{figure}
\psfig{file=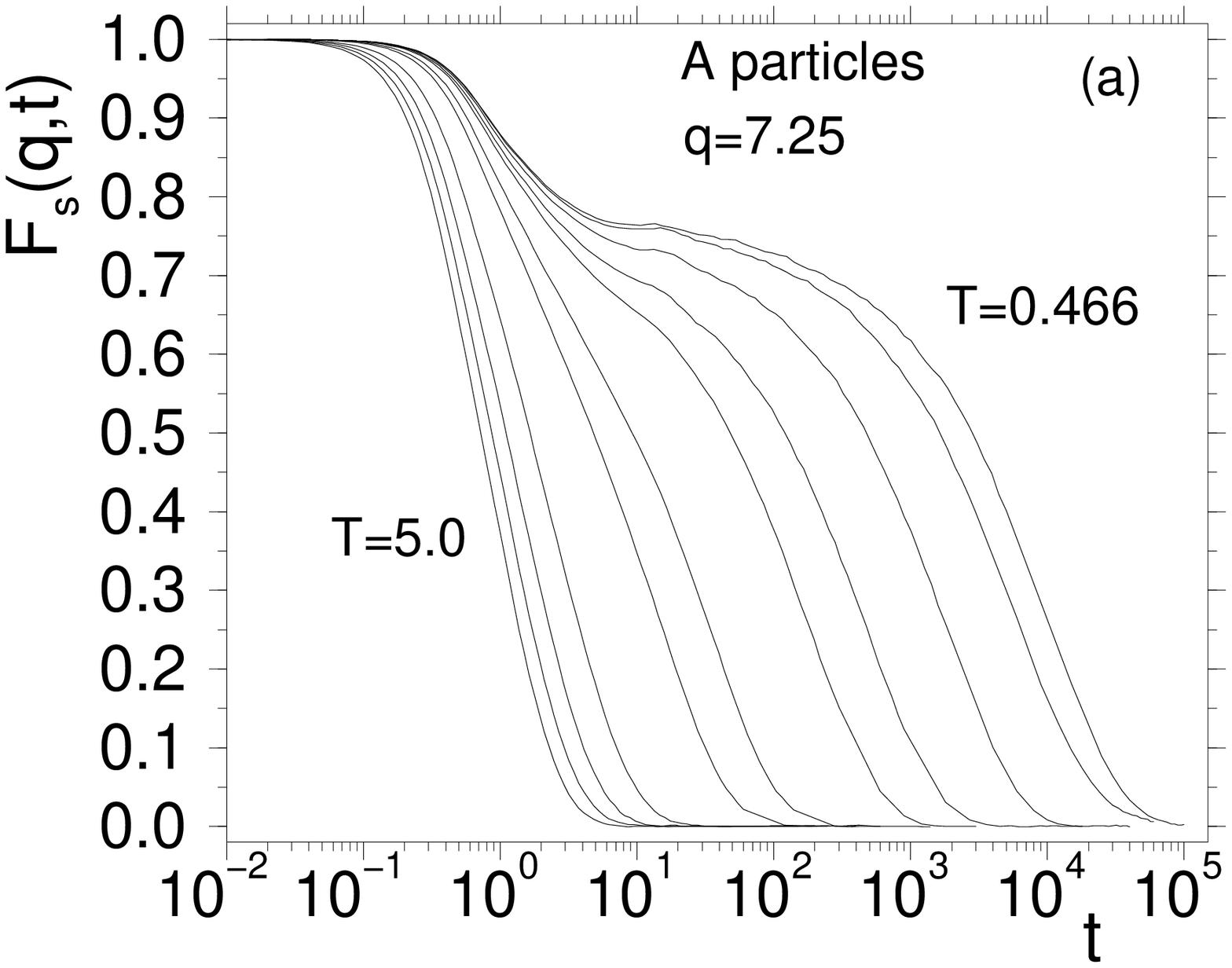,width=13cm,height=9.5cm}
\psfig{file=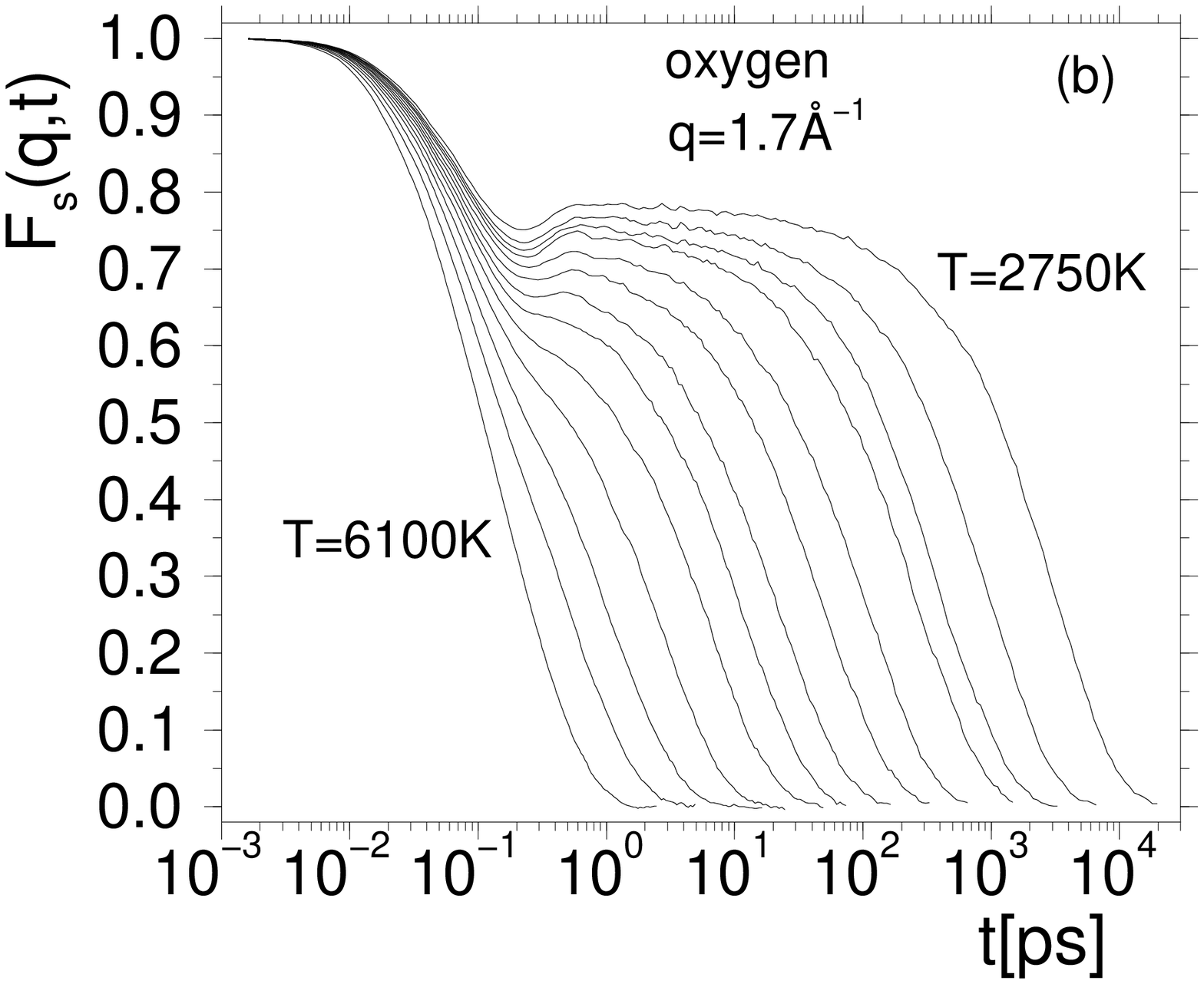,width=13cm,height=9.5cm}
\caption{Time dependence of the incoherent intermediate scattering
function for different temperatures (see Fig.~\protect\ref{fig3}). a)
Lennard-Jones system. b) Silica. From
Refs.~\protect\cite{horbach98a,kob_lj}.}
\label{fig5}
\end{figure}

In Fig.~\ref{fig5} we show the time dependence of the intermediate
scattering function for different temperatures. Let us first discuss
the relaxation dynamics for the fragile glass former,
Fig.~\ref{fig5}a.  The wave-vector $q$ corresponds to the location of
the first maximum in the structure factor (see Fig.~\ref{fig2}a) but we
have found that for the other values of $q$ a qualitatively similar
time and temperature dependence is found~\cite{kob_lj}.  After the
ballistic motion of the particles at short times, giving rise to a
quadratic dependence of $F_s(q,t)$ on time, the correlator shows at high
temperatures a crossover to an exponential decay. For low temperatures
we find a different relaxation behavior in that, after the microscopic
regime, the correlation functions show a plateau, the length of which
increases rapidly with decreasing temperatures. The reason for the
existence of this plateau is the same one we gave in the discussion of
the MSD in Fig.~\ref{fig3}, namely the cage effect, i.e.  the temporary
trapping of the particles by their neighbors. It is customary to call
the time window in which the correlator is close to the plateau the
``$\beta-$relaxation regime'' and the window in which the correlator
falls below the plateau the ``$\alpha-$relaxation regime''.

From the figure we also recognize that at low temperatures the shape of
the curves does, in the $\alpha-$relaxation regime, not depend on
temperature, an observation that we will discuss in more detail below.
Thus the whole increase of the relaxation times is due to the dynamics
in the $\beta$-relaxation regime and hence it is this regime which has
to be understood from a theoretical point of view in order to give a
correct description of the relaxation dynamics, and hence the glass
transition.

For the strong glass former the time dependence of the correlation
functions is qualitatively similar to the one of the fragile glass
former. In Fig.~\ref{fig5}b we show $F_s(q,t)$ for the oxygen atoms for
a wave-vector at the location of the first sharp diffraction peak. The
main, readily observable, difference between the relaxation behavior of
the fragile and the strong glass former is that at low temperatures the
latter shows a dip at around 0.2~ps, i.e.  shortly after the
microscopic regime and before the correlator shows the plateau. The
time at which this dip occurs corresponds to about 1.5~THz, a frequency
at which silica shows a pronounced enhancement of the density of states
over the value expected from a Debye law~\cite{bp_papers}. Therefore
this feature is called the ``boson-peak'' (``boson'' because its
temperature dependence is given by the Bose-factor). Note that this
feature is observable already at $T=3580$K, thus at temperatures far
above the (experimental) glass transition temperature of silica, which
is 1450K. (We also mention that in Ref.~\cite{horbach_diss} evidence is
given that the glass transition temperature of the BKS model is very
close to this experimental value, thus giving further support for the
reliability of this potential.) The nature of the excitations leading
to the boson-peak is still a matter of
debate~\cite{bp_papers,taraskin97,horbach98b} and we do not enter that
discussion here. (The issue is on whether the peak is due to localized
modes or due to a strong scattering of sound waves). We mention,
however, that it was found from computer simulations of systems with
different sizes, that the depth of the dip in $F_s(q,t)$ depends
strongly on the size of the system and that these finite size effects
become more severe with decreasing temperature~\cite{horbach96}. Thus
it can be concluded that the excitations that give rise to this feature
involve cooperative motion that extends at least over several nm and
that, in order to finally find the answer concerning the nature of the
peak, large systems [O($10^4$) ions] have to be analyzed so that the
mentioned finite size effects can be avoided.

\begin{figure}
\psfig{file=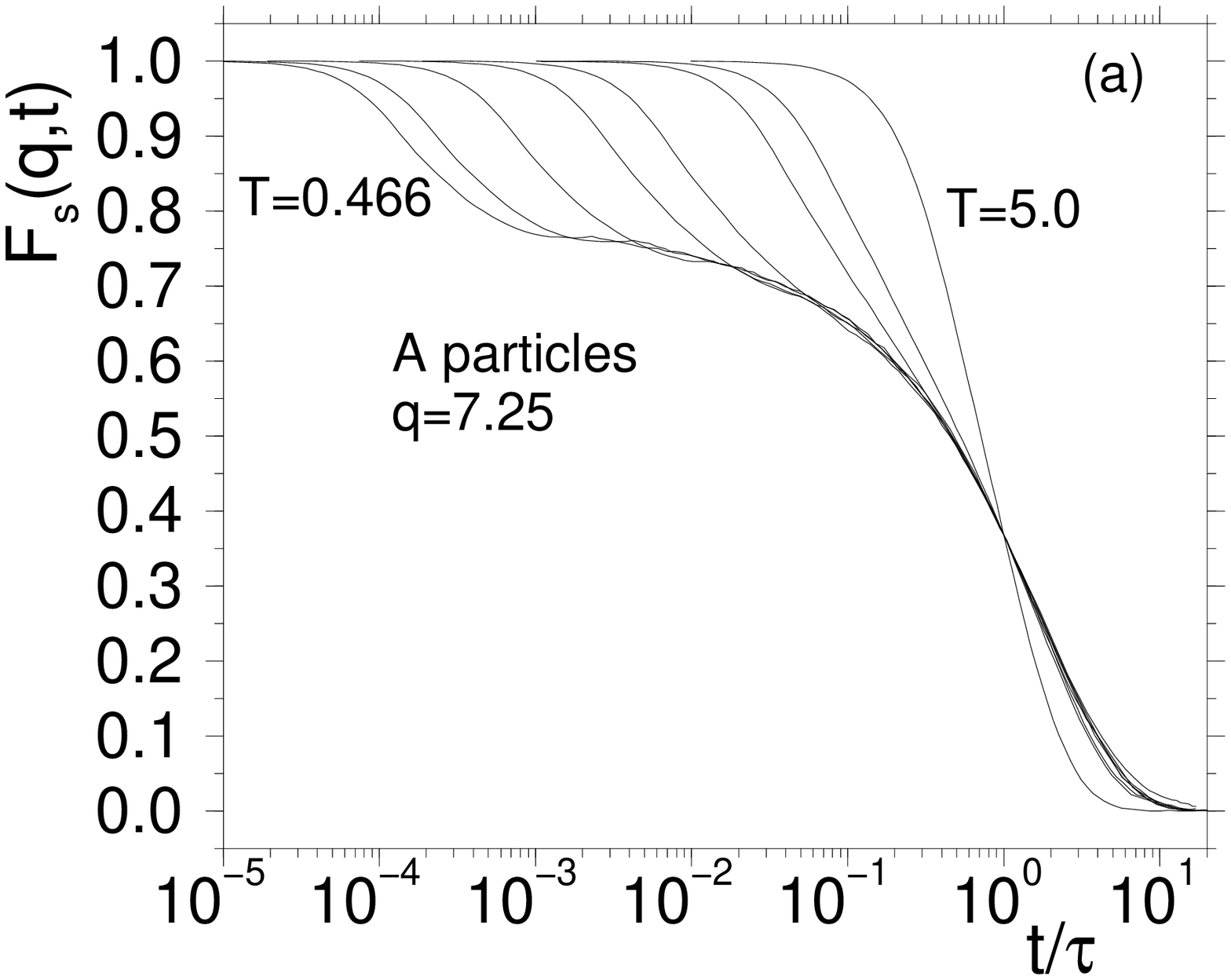,width=13cm,height=9.5cm}
\psfig{file=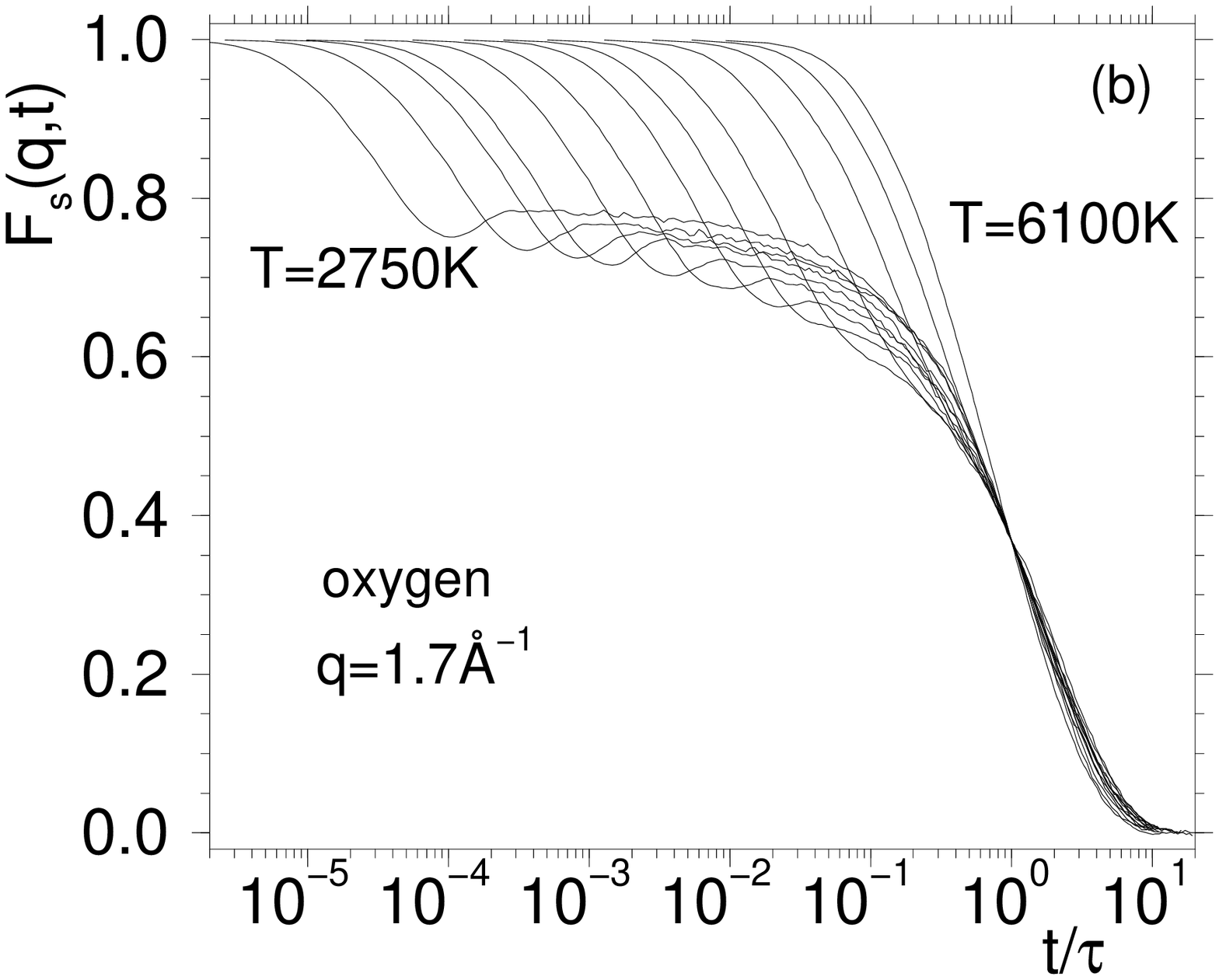,width=13cm,height=9.5cm}
\caption{Incoherent intermediate scattering function versus rescaled
time for different temperatures (see Fig.~\protect\ref{fig3}). a)
Lennard-Jones system. The curves for $T=4.0$ 3.0, and 2.0 are not
shown. b) Silica. From
Refs.~\protect\cite{horbach98a,kob_lj}.}
\label{fig6}
\end{figure}

A further difference between the relaxation behavior of strong and
fragile glass formers is concerned with the temperature dependence of
the height of the plateau in time correlation functions.  This effect
can be studied best if one plots the correlation functions versus the
rescaled time $t/\tau(T)$, where $\tau(T)$ is the $\alpha-$relaxation
time at temperature $T$.  This time can, e.g., be defined as the time
it takes a correlation function to decay to $e^{-1}$ of its initial
value. (Another possibility would be to define it as the area under the
correlator.) In Fig.~\ref{fig6} we show the so obtained figures for the
same correlators shown in Fig.~\ref{fig5}. For the fragile glass former
we find that, at low temperatures, this scaling leads to a master curve
which extends throughout the whole $\alpha-$regime, i.e. that part of
the relaxation in which the correlators fall below the plateau. Such a
master curve has been found also in other
simulations~\cite{lewis_wahnstrom,bennemann98,kammerer,hansen,wahnstrom}
and in experiments~\cite{colloids} and its existence is one of the
important predictions of MCT. The existence of this master curve is by
no means trivial, as can be recognized from the corresponding plot for
silica (Fig.~\ref{fig6}b) since for this systems the scaled curve do,
in the vicinity of the plateau, not fall at all onto a master curve. A
detailed analysis of the individual curves shows, however, that, in the
$\beta-$relaxation regime, the {\it shape} of the curves is indeed
independent of temperature~\cite{horbach_diss,horbach98c}. Thus the
only reason why they do not fall onto a master curve is the presence of
the boson-peak at short times which leads to a temperature dependent
height of the plateau. Also in experiments it has been found that in
strong glass formers the boson-peak dominates the time dependence of
the correlation function in the time range where the correlators
approach the plateau~\cite{rossler94}. Most of these findings stem,
however, from experiments in which the system is probed in the
frequency domain (such as light scattering).  Therefore such
experiments, if done at frequencies $\omega>0$, will in general not
notice whether or not the height of the plateau depends on temperature,
since such a dependency will affect only the intensity of the signal at
frequency zero. Thus the results from the computer simulation do indeed
give new insight into the dynamics of strong glass formers.

Above we have shown that for the case of silica the power-law predicted
by MCT for the temperature dependence of $D$ can be observed only for a
relatively small range of the diffusion constant.  From
Fig.~\ref{fig6}b one sees that, due to the strong influence of the
boson-peak in the $\beta$-relaxation regime, also the predicted master
curve is not observed. Therefore one {\it might} be tempted to argue
that MCT is not a very useful description of the dynamics of SiO$_2$.
This point of view is, however, far too pessimistic, since there are
predictions of the theory which are valid also in the presence of the
mentioned hopping processes, such as, e.g., the so-called factorization
property, which states that in the $beta$-relaxation regime any time
correlation function $\Phi(t)$ can be written as $\Phi(t)=f+hG(t)$,
where $G(t)$ is a system universal function and $f$ and $h$ will depend
on the correlation function $\Phi$. It has been
shown~\cite{horbach_diss,horbach98c} that for the case of silica this
factorization property holds very well, thus showing that MCT is able
to make relevant (and correct) predictions on the dynamics for this
(strong) type of glass former as well.

Having now some understanding of the temperature dependence of the
relaxation dynamics of the Lennard-Jones and the silica system we can
compare it with that of other glass-formers. Sciortino {\it et al.} have
done extensive simulations of the dynamics of supercooled
water~\cite{sciortino}, a system which shares many structural and
thermodynamical properties with
silica~\cite{debenedetti97,poole97,vessal89,speedy97}. These simulations
have shown that in the temperature range accessible to equilibrium
simulations, the diffusion constant shows a power-law
dependence~\cite{sciortino}, a result that is supported also by
experiments~\cite{debenedetti97,water_pow_law_exp}. Thus, although the
system is {\it structurally} much more similar to the strong glass
former silica, it behaves {\it dynamically} like a fragile glass
former. From our findings for silica we therefore conclude that the strong
hopping processes found in silica are less pronounced in water than in
silica (which might, however, also have to do with the fact that the
water simulations have been done with H$_2$O molecules that could not
dissociate).

For the early $\beta-$relaxation regime it is found~\cite{sciortino}
that also water shows a small dip, as we showed it to be present in
silica (see Fig.~\ref{fig6}b). However, contrary to the case of silica,
this dip does not destroy the master curve in the late $\beta$-relaxation
regime when the correlators are plotted versus $t/\tau(T)$. Thus the
dynamics of water seems to behave in certain aspects like a fragile
glass former and in other aspects like a strong one. This duality is
most likely due to the network structure since the simple atomic and
molecular liquids, that do not have the tendency to form a network,
have been found to show essentially the same dynamics as the
Lennard-Jones system discussed
here~\cite{lewis_wahnstrom,kammerer,hansen,wahnstrom,teichler96}.  Also
the dynamics of polymeric glass formers is in many aspects quite
similar to the one of simple liquids~\cite{bennemann98}. However,
because of the length of the molecules these systems do also show
interesting dynamic effects which are not present in simple liquids,
such as the motion of side chains or reptative movements. For a more
thorough discussion of these issues we refer the reader to the review
articles by Clarke~\cite{clarke95} and Paul and
Baschnagel~\cite{paul95}.

The results presented so far are mainly useful to understand the
slowing down of the relaxation dynamics on a {\it qualitative} level.
However, it is also possible to use computer simulations to study this
relaxation dynamics on a {\it quantitative} level and thus to test
predictions of theories that attempt to describe this dynamics. In
the following we will therefore briefly describe a few simulations
which have been done to test the validity of such theories.

\begin{figure}
\psfig{file=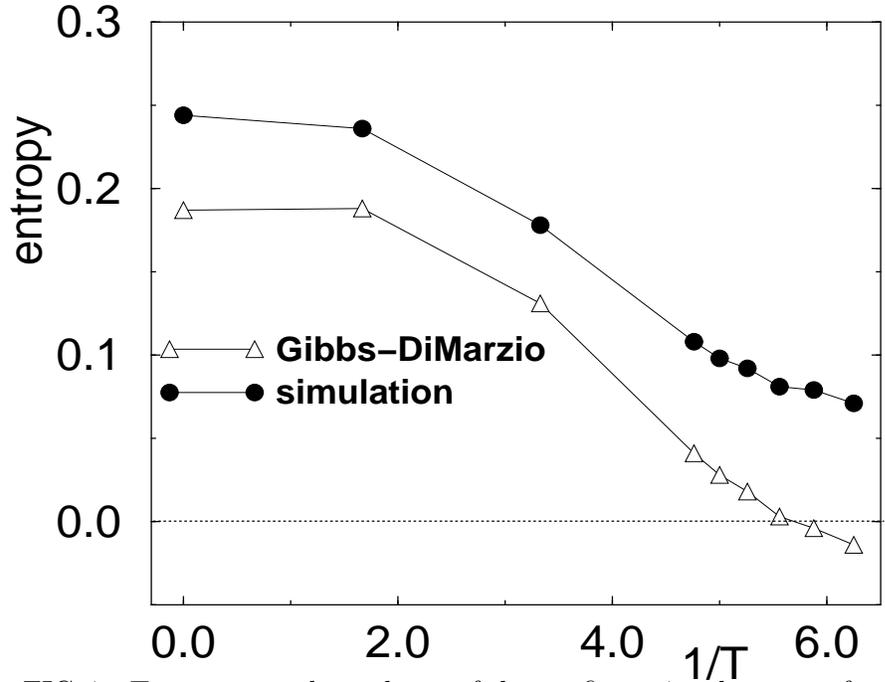,width=12cm,height=9.0cm}
\caption{Temperature dependence of the configurational entropy for a
polymer model as determined from a Monte Carlo simulation (filled
symbols) and from the prediction of the theory by Gibbs and DiMarzio
(open symbols).  Adapted from Ref.~\protect\cite{wolfgardt96}.}
\label{fig7}
\end{figure}

As already mentioned in Section~\ref{sec1}, two very prominent theories
are the ones of Gibbs and DiMarzio~\cite{gibbs_dimarzio58}, which is
very popular in the community of polymer scientists, and the so-called
mode-coupling theory which seems to be applicable for a very large
variety of glass-formers~\cite{mct}. The basic idea of theses two
theories is very different. In the Gibbs-DiMarzio approach one starts
from polymers which are placed on a lattice. By counting the number of
ways the polymers of the melt can be placed on the lattice, Gibbs and
DiMarzio found that there exists a critical packing fraction above
which this configurational entropy goes to zero in a continous way.
From this they concluded the existence of a second order phase
transition at this point, which is acompanied by the usual critical
slowing down of the dynamics. This slowing down then gives rise to the
glass transition. It has been found that this theory is able, e.g., to
rationalize the dependence of the glass transition temperature on the
length of the polymers, on the concentration of plastisizer, and other
relevant quantities~\cite{dimarzio97}. It was therefore quite
surprising when a few years ago a simulation of Wolfgardt {\it et al.}
showed that the observed slowing down of the relaxation dynamics of a
dense polymer melt is {\it not} due to the fact that the
configurational entropy of the system becomes zero~\cite{wolfgardt96}.
This can be seen in Fig.~\ref{fig7} in which the entropy is plotted
versus the inverse temperature. The open symbols correspond to the
entropy as given by the (approximate) expression proposed by Gibbs and
DiMarzio and one sees that this prediction gives a zero value of the
entropy at a finite temperature.  (It has to be emphasized that the
quantities entering the Gibbs-DiMarzio expression have been calculated
from the simulation as well and thus no approximation on that level
is made.) The filled symbols are the real values of the entropy of the
system which have been measured by thermodynamic integration. From the
figure it becomes clear that this real entropy does not go to zero in
the temperature range in which the Gibbs-DiMarzio expression becomes
zero, that, however, it shows a noticeable decrease in this temperature
range. Thus one has to conclude that the success of the Gibbs-DiMarzio
theory relies probably on the fact that the real entropy decreases
significantly in the vicinity of the critical temperature, that the
theory gives a quite realistic description of this decrease {\it on the
qualitative level}, and that most experiments measure only quantities
which are related to the various {\it derivatives} of the entropy. 

The second theory that we will discuss here to some extend, the
mode-coupling theory, has a very different explanation for the slowing
down of the system upon cooling. From the theory of the dynamics of
simple liquids in the vicinity of the triple point it is known that the
equations of motion for density correlation functions have nonlinear
terms which are needed to describe back-flow effects and cage
effects~\cite{sjogren80} and MCT is such a set of equations of motion.
It is found that with decreasing temperature the nonlinear terms become
stronger and will lead to a dynamical feedback effect that slows down the
relaxation of the density correlators.  Within a certain approximation
to these equations, leading to what is known as the ``ideal MCT'', this
feedback mechanism becomes so strong that at a certain temperature
$T_c$ the correlation functions do not decay to zero anymore, i.e. the
system has undergone a glass transition. Using this $T_c$ as a
reference temperature, the theory makes detailed predictions regarding
the time and temperature dependence of correlation functions, such as
the intermediate scattering function discussed above.

One of the important aspects of the theory is that the equations of
motion for the density correlators depend only on {\it static}
quantities, such as the structure factor. Since such quantities can be
obtained quite easily with high precision from experiments or
simulations, it is thus possible to measure the static quantities,
solve the MCT equations and compare the so obtained theoretical
curves for the time dependent correlation functions with the one
measured in the experiment or in the computer simulation. Therefore very
stringent tests of the theory become possible on a qualitative as
well as {\it quantitative} level.

The outcome of such a test has recently been reported by Nauroth {\it
et al.}~\cite{nauroth97}. In that work the MCT equations for the binary
Lennard-Jones system discussed above have been solved numerically and
their solutions at long times compared with the results from
simulations. Among the quantities investigated were the wave-vector
dependence of the height of the plateau in the intermediate scattering
function (see Fig.~\ref{fig5}) at the critical temperature $T_c$, a
quantity which is also called the Edwards-Anderson, or nonergodicity
parameter $f_c(q)$. 

\begin{figure}
\psfig{file=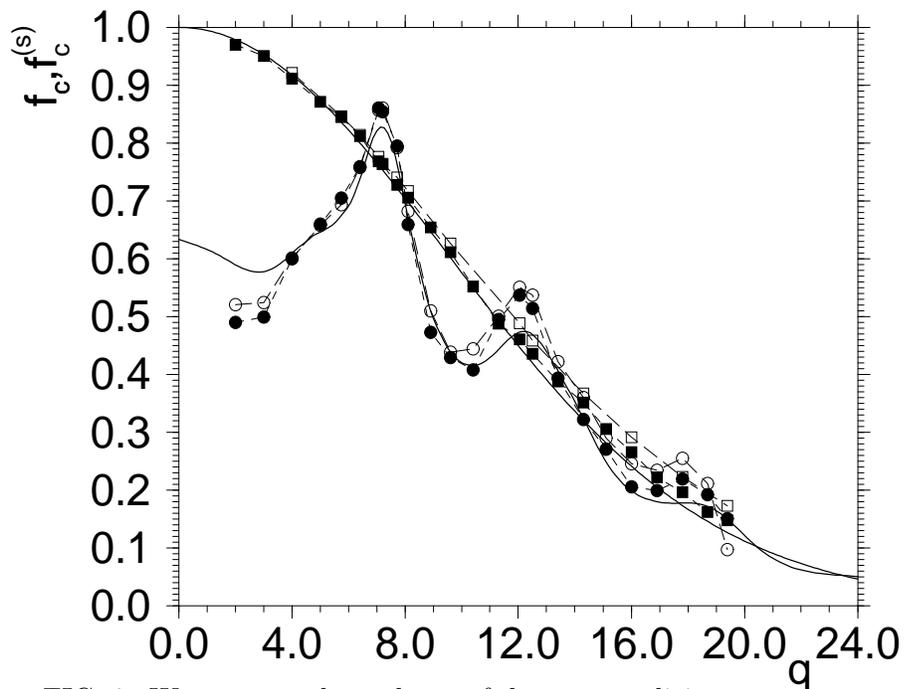,width=12cm,height=9.0cm}
\caption{Wave-vector dependence of the nonergodicity parameter as
predicted from MCT (solid lines), as measures in a Lennard-Jones
system with Newtonian dynamics (open symbols) and a stochastic
dynamics (filled symbols). The Gaussian shaped curves are for the
incoherent intermediate scattering function and the oscillatory
curves are for the coherent intermediate scattering function. From
Ref.~\protect\cite{gleim98}.}
\label{fig8}
\end{figure}

In Fig.~\ref{fig8} we show $f_c(q)$ as predicted by MCT (solid curves)
and the one from the computer simulation (open symbols). The Gaussian
shaped curve is the nonergodicity parameter for the incoherent
intermediate scattering function $F_s$ of the $A$ particles
[Eq.~(\ref{eq3})] and the oscillatory curves are for the coherent
intermediate scattering function for the $A-A$
correlation~\cite{theory_liquids}. We emphasize that for the
calculation of the theoretical curves no fit parameter of any kind was
used. The only input was the temperature dependence of the partial
structure factors, which were determined from the simulation.
From the figure we recognize that the theory gives
an excellent description of the wave-vector dependence of $f_c$ and we
thus conclude that the theory is indeed able to make precise {\it
quantitative} predictions on the dynamics of supercooled simple
liquids.

Also included in the figure are the results of a computer simulation of
the mentioned binary Lennard-Jones system, but instead of the Newtonian
dynamics used in Ref.~\cite{kob_lj} a Brownian-like (stochastic)
dynamics was used~\cite{gleim98}.  In this type of dynamics the
equations of motion of the particles are given by

\begin{equation}
m\ddot{\bf r}_j+\sum_{i\neq j} 
\frac{\partial V(|{\bf r}_i-{\bf r}_j|)}{\partial {\bf r}_j} +\zeta
\dot{\bf r}_j+\eta_j(t)=0 \quad ,
\label{eq4}
\end{equation}
where $\zeta$ is a friction constant whose value is related to the
amplitude of the white noise $\eta_j(t)$ by the fluctuation dissipation
theorem, i.e. $\langle \eta_j(t)\eta_i(t')\rangle=
6k_BT\delta(t-t')\delta_{ji}$. There are two reasons for studying a
system with such a stochastic dynamics. Firstly the system with these
stochastic forces will probably have a dynamics which is quite similar
to the one of a colloidal particle in a suspension, a model system for
which beautiful light scattering experiments have been performed to
study the glass transition~\cite{colloids}. Thus it is interesting to
see whether the relaxation dynamics of the Lennard-Jones system is
indeed very similar to the one found in these experiments.  The second
reason is related to a prediction of MCT which states that at long
times the relaxation dynamics is independent of the microscopic
dynamics. Thus we can compare the relaxation behavior of the system
with the stochastic dynamics with the one of the Newtonian dynamics and
see in what aspects the two differ. Work in this direction has already
been done by L\"owen {\it et al.}~\cite{lowen91}. These authors made a
simulation of a supercooled polydisperse mixture of particles
interacting with a Yukawa potential and compared the relaxation
dynamics of this system with a Newtonian dynamics to the one with a
Brownian dynamics.  It was found that in the $\beta-$relaxation regime
the dynamics depends on the microscopic dynamics, that, however, the
$\alpha-$relaxation was independent of it. Qualitatively similar
results have been found by Gleim {\it et al.} in the mentioned
simulation of the Lennard-Jones system~\cite{gleim98,gleim_diss}. In
that work it was shown that at low temperatures the whole relaxation
dynamics is independent of the microscopic dynamics, if one leaves
aside the relaxation at very short (microscopic) times. An example for
this finding is shown in Fig.~\ref{fig8}, were we have also included
the wave-vector dependence of the nonergodicity parameter for the
system with the stochastic dynamics (filled symbols). We see that this
$q$-dependence is essentially independent of the microscopic dynamics,
thus confirming this prediction of the theory. Furthermore we also
recognize that essentially for all wave-vectors the curves for the
stochastic dynamics agree even better with the theoretical ones than
the curves for the Newtonian dynamics do. The reason for this might be
that MCT assumes that the time scale of the $\beta-$relaxation, at
which the height of the plateau is measured, is separated well from the
one of the microscopic relaxation.  Since in the stochastic dynamics
the phonons are strongly damped it can be expected that {\it
effectively} this separation is larger in the case of the stochastic
dynamics than for the case of the Newtonian dynamics, where no damping
is present, and that therefore the assumption of the theory is better
fulfilled.

\subsection{Dynamics below the glass transition temperature}
\label{sec3b}

In the previous subsection we have discussed the equilibrium relaxation
dynamics of glass-forming liquids in their supercooled state, i.e. at
temperatures above the glass transition temperature $T_g$. Since we
have seen that with decreasing temperature the relaxation time $\tau$
increases quickly, we expect that there will exist a temperature $T_g$
at which $\tau$ will exceed the time scale of the experiment or
simulation and hence it will no longer be possible to equilibrate the
system. One might guess that at temperatures below $T_g$ the system is
essentially frozen, i.e. that its dynamics is similar to the
vibrational dynamics of a solid. To a certain extend this view is
certainly correct and solid state concepts, like, e.g., the density of
states, have successfully been applied to describe the dynamics of
disordered systems or to compute their specific
heat~\cite{horbach98d}.  However, as we will show in this section, even
at low temperatures the dynamics of disordered systems is not purely of
vibrational type but has also a very interesting relaxational
component. This slow relaxation is known as ``aging'' and leads to a
slow time dependence of various material properties, such as
brittleness, density, etc.  The importance of such aging phenomena has
been realized for quite a long time~\cite{struik78_mckenna89} (and they
have been described by means of phenomenological theories) but it is
only in recent years that sound theoretical concepts have been
developed in order to describe this type of
dynamics~\cite{aging_theo_rev,aging_theo,nieuwenhuizen98,domain_growth}.
Apart from extensive investigations on polymeric
systems~\cite{struik78_mckenna89}, aging phenomena have so far mainly
been investigated in spin glasses by means of experiments and computer
simulations~\cite{rieger95,aging_theo_rev,aging_exp_spin,aging_sim_spin}.
Only very recently experiments and computer simulations of structural
glasses and other disordered systems have been made in order to
investigate these
phenomena~\cite{parisi97,aging_sim,aging_rest,aging_exp_glass,kob97,barrat98,andrejew96,kob98}
so that the validity of the various theoretical approaches can be
tested. The results of these investigations do not yet allow to decide
which one of the theoretical pictures proposed (``droplet model'',
``trap model'',
mean-field)~\cite{aging_theo_rev,aging_theo,nieuwenhuizen98,domain_growth,aging_rest}
is appropriate to describe the aging dynamics of structural glasses but
it can be expected that further studies will ultimatively rule out
certain scenarios. Therefore the goal of the following presentation of
the results of such investigations is not to advocate any particular
theory but rather to familiarize the reader with the occurring effects
and to show how computer simulations might help to decide which
theoretical picture is appropriate to describe this aging dynamics.

\begin{figure}
\psfig{file=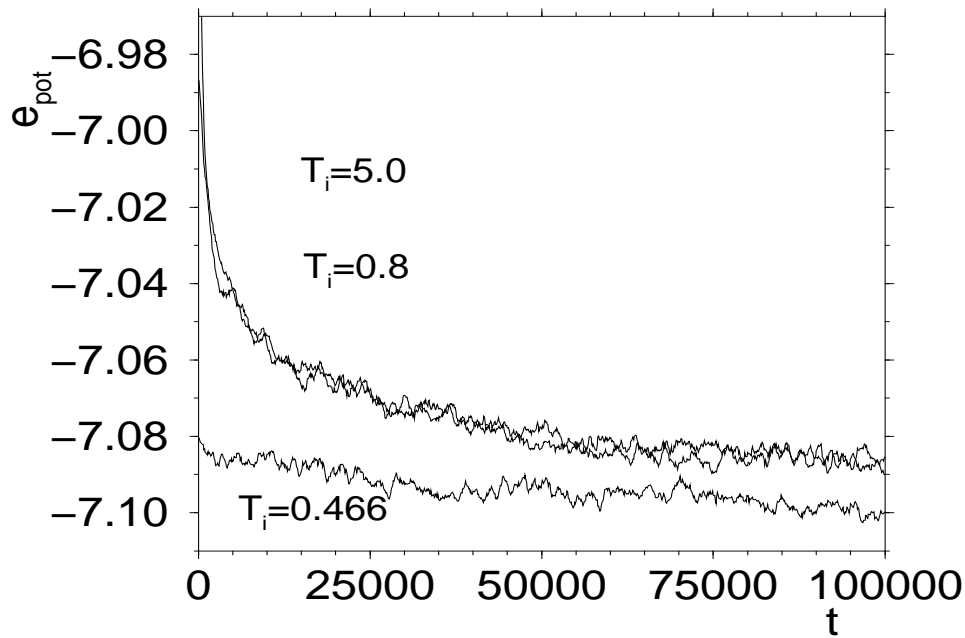,width=13cm,height=9.5cm}
\caption{Time dependence of the potential energy of a Lennard-Jones
system which at time zero has been quenched from an initial temperature
$T_i$ to a final temperature $T_f=0.4$. Adapted from
Ref.~\protect\cite{kob97}.}
\label{fig9}
\end{figure}

In order to study aging effects it is of course necessary to first drive the
system out of equilibrium. In structural glasses this can be done, e.g,
by decreasing the temperature below the glass transition temperature or
by compressing the system beyond a ``critical'' density. If the out of
equilibrium situation is generated by a temperature jump, one proceeds
as follows. The system is first equilibrated at an initial temperature
$T_i$. At time $t=0$ the system is quenched to a temperature $T_f$
which is significantly lower than the glass transition temperature
$T_g$, where $T_g$ is given by the time scale of the experiment. After
the quench the system is allowed to evolve for a waiting time $t_w$,
after which the measurements start. The relaxation of the system is now
studied as a function of $t_w$ and $\tau$,the time elapsed since the
start of the measurement, i.e. since $t=t_w$. In a recent computer
simulation Kob and Barrat studied the relaxation dynamics of the binary
Lennard-Jones system discussed above after such a quench~\cite{kob97}. In
Fig.~\ref{fig9} we show the time dependence of $e_{pot}$, the potential
energy of the system, after such a quench to $T_f=0.4$ for three
different initial temperatures $T_i=5.0$, 0.8 and 0.466. Note that the
glass transition temperature of the system for very long simulations
(O$(10^8)$ time steps) is around $T_c=0.435$~\cite{kob_lj}.  One
recognizes that, after a relatively fast decay, the potential energy
for the two larger values of $T_i$ is almost constant and also the
curve for the lowest $T_i$ depends only weakly on time. A more
careful analysis of this time dependence shows that it is approximated
well by a power-law with a small exponent around 0.14, or by a
logarithmic time dependence.  Since this dependence is so weak one
therefore might erroneously conclude that at the end of the run the
system has equilibrated, or is at least quite close to equilibrium. We
know, however, that this is clearly not the case, since we are at a
temperature at which the relaxation time of the system exceeds by
several orders of magnitude the time scale of the simulation.  Hence
this should be taken as a warning in using a one-time quantity as a
probe of whether or not the system has been equilibrated or not. (By
``one-time quantity'' we mean an observable which {\it in equilibrium}
is a constant, such as the energy, pressure or density.) Thus we
conclude that the investigation of such quantities is not very well
suited to study aging phenomena, a result in agreement with simulations
by Andrejew and Baschnagel of aging effects in a polymer
melt~\cite{andrejew96}.

\begin{figure}
\psfig{file=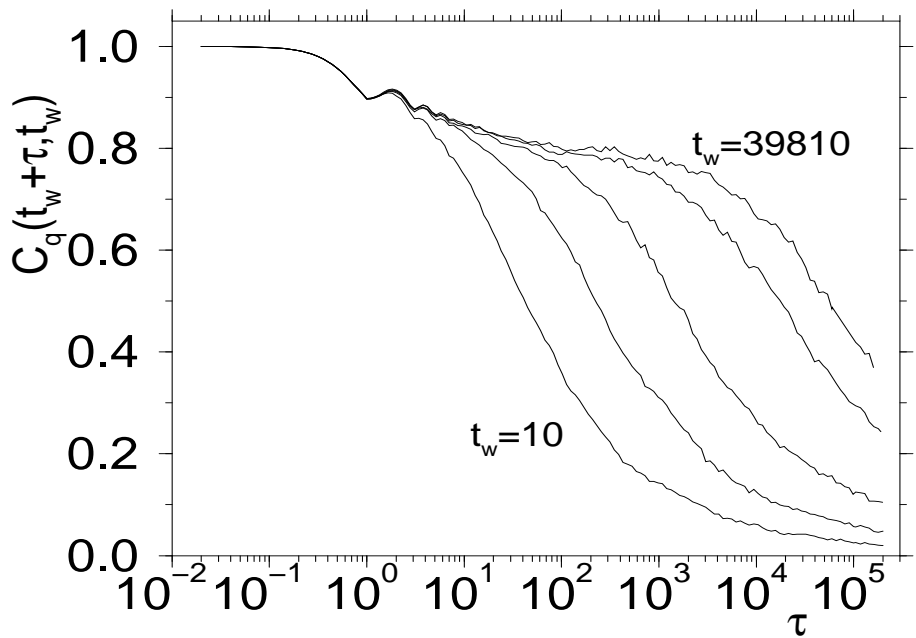,width=13cm,height=9.5cm}
\caption{Time dependence of the out of equilibrium correlation function
$C_q(t_w+\tau,t_w)$ for different waiting times $t_w$. From left to
right: $t_w=10$, 100, 1000, 10000 and 39810. The wave-vector $q$ is
7.25, the location of the maximum in the static structure factor.
Adapted from Ref.~\protect\cite{kob97}.}
\label{fig10}
\end{figure}

Much more sensitive quantities to study the aging dynamics are
``two-time'' correlation functions, such as the generalization of the
intermediate scattering function $F_s(q,t)$ [Eq.~(\ref{eq3})] to a
non-equilibrium situation. We recall that by definition of
equilibrium, a time correlation function depends only on {\it time
differences}, i.e.
\begin{equation}
F_s(q,\tau)=\frac{1}{N} \sum_{j=1}^N\langle \exp(i {\bf q} \cdot [{\bf
r}_j(\tau)-{\bf r}_j(0)] \rangle= 
\frac{1}{N} \sum_{j=1}^N\langle \exp(i {\bf q} \cdot [{\bf
r}_j(t_w+\tau)-{\bf r}_j(t_w)] \rangle\quad .
\label{eq5}
\end{equation}
Since the quench breaks the time translation invariance of the
system, the second equality no longer holds and we therefore define
the out of equilibrium time correlation function
\begin{equation}
C_q(t_w+\tau,t_w) = \frac{1}{N} \sum_{j=1}^N\langle \exp(i {\bf q}
\cdot [{\bf r}_j(t_w+\tau)-{\bf r}_j(t_w)] \rangle\quad .
\label{eq6}
\end{equation}
Here $\langle . \rangle$ stands for the average over the temperature
{\it before} the quench.  The time dependence of $C_q(t_w+\tau,t_w)$
for different waiting times is shown in Fig.~\ref{fig10}. We see that
this two-time correlation function shows a very strong dependence on
the waiting time and is therefore well suited to study the aging
dynamics. For short $t_w$ the function decays very quickly, since the
typical configuration of the particles at $T_f$ is very different from
the one at $T_i$. Therefore the system is subject to a large driving
force which will move it towards a part of phase space which is more
typical for the temperature $T_f$ and thus the system will quickly
decorrelate from the initial configuration.  If the waiting time is
increased, the driving force is smaller and at short and intermediate
times $\tau$ the system will explore only that part of phase space
which corresponds to the vibrational motion of the particles in their
cages. This is the reason why the correlation functions show a plateau
at intermediate times and look qualitatively quite similar to the ones
in equilibrium (cf.  Fig.~\ref{fig5}a).  Only for larger times $\tau$
does the correlation function decay to zero and the time at which this
final decay is observed increases with $t_w$. Although the
off-equilibrium relaxation curves $C_q(t_w+\tau,t_w)$ for large $t_w$
look qualitatively similar to the ones in equilibrium [i.e. to
$F_s(q,t)$] a detailed analysis shows that there are important
differences. E.g. it is well known that at long times the equilibrium
relaxation curves of glass forming liquids can be fitted very well by a
Kohlrausch-Williams-Watts function, $A\exp(-(t/\tau_{KWW})^{\beta})$,
and this is also the case for the present Lennard-Jones
system~\cite{kob_lj}. For the out of equilibrium function it is,
however, found that the relaxation at long times is given by a power
law with an exponent around 0.4~\cite{kob98} a result which is in
qualitative agreement with results of simulations of spin
glasses~\cite{rieger95,aging_sim_spin}. Hence we conclude that despite
the apparent similarity of the curves for equilibrium and out of
equilibrium relaxation, the two types of functions differ
significantly.  Therefore it is not appropriate to use the latter ones
as an approximation for the former ones, as it is sometimes done when
time correlation functions for temperatures above and below $T_g$ (as
determined from the simulation) are mixed together in the analysis of
the data.

In the discussion of Fig.~\ref{fig9} we concluded that a one-time
quantity is not a good indicator for whether or not the system has
reached equilibrium.  Often it is argued that the decay of a time
correlation function, or seeing a diffusive (i.e. linear) time
dependence of the mean squared displacement within the time span of the
simulation, is a sufficient condition for having reached equilibrium.
From Fig.~\ref{fig10} one recognizes however, that this is not the case
at all, since the correlation functions decay also in the
non-equilibrium situation, a result which was also demonstrated nicely
by Baschnagel~\cite{baschnagel94}, and in a recent simulation of a soft
sphere system Parisi demonstrated that a linear dependence of the MSD
does not imply at all that one has reached
equilibrium~\cite{parisi97}.

One of the interesting results of the theories of aging concerns the
violation of the fluctuation-dissipation theorem
(FDT)~\cite{aging_theo_rev,nieuwenhuizen98} which in equilibrium
relates $R_A(t)$, the response of an observable $A$ to its conjugate
field, to the time autocorrelation function $C_A(t)=\langle A(t)A(0)
\rangle$, i.e.  $R_A(t)=-(k_BT)^{-1} \partial C_A(t)/\partial t$. For
the out of equilibrium case this relation is no longer valid and it is
generalized to
\begin{equation}
R_A(t',t)=\frac{1}{k_BT}X_A(t',t) \frac{\partial C_A(t',t)}{\partial t}
\quad,
\label{eq7}
\end{equation}
where $t'\geq t$ and the quantity $X_A(t',t)\leq 1$ is defined by this
equation. From this relation it becomes clear that the quantity
$T/X_A(t',t)$ can be viewed as the effective temperature for which the
FTD holds~\cite{cugliandolo97} (see also Ref.~\cite{nieuwenhuizen98}).
Note that in the field of glass science the concept of a ``fictive''
temperature has been introduced long ago by Tool and
Eichlin~\cite{tool31} but, to our knowledge, has never been based on a
solid statistical mechanics foundation. In contrast to this the
definition of such a temperature via Eq.~(\ref{eq7}) does not have this
drawback and permits to measure this temperature in experiments or
simulations. From the equation it also becomes immediately obvious
that the ratio $T/X_A(t',t)$ will in general depend on time and on the
observable considered.

If the observable of interest is the coherent intermediate scattering
function $C_q(t_w+\tau,t_w)$ one can measure the response
$R(t_w+\tau,t_w)$ (in principle) as follows~\footnote{More details on
this calculation can be found in Ref.~\cite{barrat98}.}. After the
waiting time $t_w$, a space dependent sinusoidal field with wave-vector
$q$, which couples to the particle density and has an amplitude $V_0$,
is turned on and the resulting change in the density distribution for
wave-vector $q$ is measured. After a measuring time $\tau$ one
therefore obtains the integrated response $M(t_w+\tau,t_w)$ given by

\begin{equation}
V_0 M(t_w+\tau,t_w)  =  V_0 \int_{t_w}^{t_w+\tau} R(t_w+\tau,t) dt \quad .
\label{eq8}
\end{equation}

For large values of $\tau$ and $t_w$ it is expected that the
FDT-violation factor $X(t_w+\tau,t_w)$ in Eq.~(\ref{eq7}) becomes a
function of $C(t_w+\tau,t_w)$ only, i.e.
$X(t_w+\tau,t_w)=x(C(t_w+\tau,t_w))$, where $x$ is a function of {\it
one} variable~\cite{aging_theo_rev}. From this relation and
Eq.~(\ref{eq8}) one thus derives
\begin{equation}
M(C_q)=\frac{1}{k_BT}\int_{C_q}^1 x(c) dc,
\label{eq9}
\end{equation}
where we used the fact that $C_q(t_w+\tau,t_w)=1$ for $\tau=0$. This
equation thus says that a parametric plot of $k_BTM$ versus $C_q$, with
time $\tau$ as a parameter, will give us information on the function
$x(c)$ and hence on the FDT-violation factor $X(t_w+\tau,t_w)$.

\begin{figure}
\psfig{file=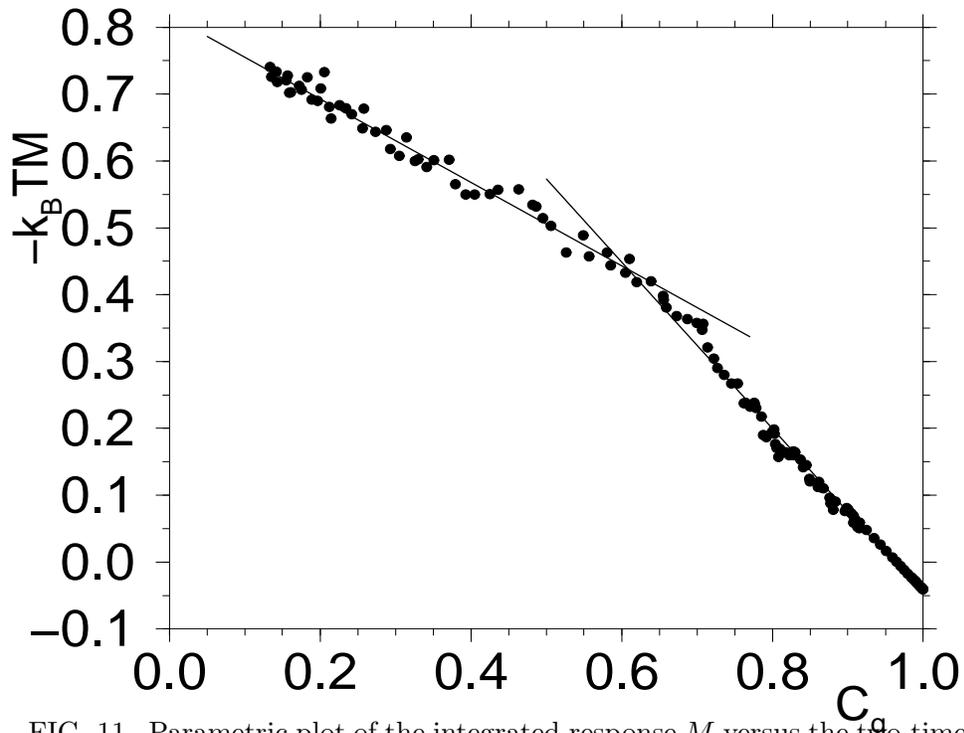,width=13cm,height=9.5cm}
\caption{Parametric plot of the integrated response $M$ versus the
two-time correlation function $C_q$. The two straight lines have
slopes around $-0.6$ and $-1$, respectively. Adapted from
Ref.~\protect\cite{kob98}.}
\label{fig11}
\end{figure}

In Fig.~\ref{fig11} we show such a parametric plot for the same
correlation function shown in Fig.~\ref{fig10} and the corresponding
integrated response $M$. The values of $t_w$ and $T_f$ are 1000 and
0.4, respectively. We see that for short times $\tau$, corresponding to
large values of $C_q$, the data points can be approximated well by a
straight line with a slope around $-1$ (see figure). This means that
the FDT-violation factor $X$ is $1$, i.e.  that the FDT holds and the
system behaves like in equilibrium. However, for larger times
(corresponding to smaller values of $C_q$) the FDT is violated since
the data points do not fall anymore on a straight line with slope
$-1$.  What is observed instead is a straight line with a slope around
$-0.62$.  This means that in this time regime $X(t_w+\tau,t_w)$ is
constant and has a value around 0.62, which corresponds to an effective
temperature of $0.4/0.62 \approx 0.64$.  Qualitatively similar results
have been found in the seminal work of Parisi in which the violation of
the FDT has been investigated for a soft sphere system~\cite{parisi97}.
In that simulation the observable of interest was the mean squared
displacement and it was shown that the FDT violation factor $X$ shows a
linear dependence on the final temperature $T_f$ of the quench, a
result which has been confirmed also by Barrat and
Kob~\cite{barrat98}.

Note that the fact that in the non-FDT region we find a straight line
with a {\it finite} slope is not trivial at all, since certain theories of
aging, such as domain growth~\cite{domain_growth}, predict a straight
line with slope zero and others a parabola-like
dependence~\cite{aging_theo_rev}. Hence we see that these type of
measurements can indeed be used to collect evidence for or against a
theoretical scenario. 

\section{Conclusion}
\label{sec4}
In this review we have discussed some results of computer simulations
of supercooled liquids. By now the literature on this topic is so vast
that a comprehensive review is unfortunately no longer possible and
thus we have focussed on only a few topics, namely on the (metastable)
equilibrium dynamics above $T_g$ for a strong and fragile glass former
and the non-equilibrium dynamics below $T_g$. Whereas the investigation
of the former type of question is today still dominated by real
experimentalist, the latter seems to have been studied in much more
detail by means of computer simulations, since for the moment they are
better adapted to investigate such problems.  We hope however that in
the near future this situation changes, since such investigations allow
to learn more about the structure of the phase space of the system. (We
mention that in the past some properties of phase space have been
investigated by determining the inherent structure of liquids and
studying normal modes~\cite{inher_str_norm_modes}.) If the structure of
this phase space is understood well, e.g.  whether or not it is
organized in a hierarchical way, it might become possible to understand
the connection between systems in which the disorder is quenched (such
as spin glasses) and systems with self generated disorder (e.g.
structural glasses).

Other types of questions in which computer simulations are probably
very valuable, are to investigate the {\it equilibrium} dynamics of
supercooled liquids in the region between the MCT temperature $T_c$ and
the experimental glass transition $T_g$ on a microscopic level.
Although for fragile glass formers such simulations are currently not
quite feasible, they will be possible in a few years. Since at the
moment there is no complete theory for the dynamics of liquids in this
temperature range and real experiments rarely allow to probe the system
on the microscopic level in sufficient detail, such simulations will be
an excellent tool to gain insight into this question and thus possibly
serve as a guide to the development of reliable theories.

Apart from these types of simulations which are motivated mainly by the
wish to have a sound theoretical understanding of glassy materials,
there are of course also those simulations which are done to calculate
properties of specific materials. For example it is possible to predict
the temperature dependence of the specific heat of silica, for
temperatures above 100~K up to $T_g$, to within a few percent of the
experimental values~\cite{horbach98d}, or the dynamic structure factor
with quite high accuracy~\cite{taraskin97,horbach_diss,horbach98c}. It
can be expected that in a few years the quality of the available
potentials will increase even further and that soon also potentials of
a bit more exotic materials (such as multicomponent systems) will be
determined.  This will then in turn open the door to many types of
simulations related more closely to materials science and thus allow to
use all the know-how and insight gained in the investigation of less
complex models also in technological more relevant materials.

Acknowledgements: I thank H. C. Andersen, C. A. Angell, J.-L. Barrat,
K. Binder, T.  Gleim, J. Horbach, M. Nauroth, and K. Vollmayr with whom I
had the pleasure to collaborate on some of the work that was presented
here, and W. Paul for useful comments on the manuscript. I am also
deeply grateful to all the people who have taught me most of the things
I know on this subject and whose shear number regretfully prevents me
to list them here by name. Part of this work was supported by the
Deutsche Forschungsgemeinschaft under SFB 262, the Pole Scientifique
de Mod\'elisation Num\'erique at the ENS-Lyon and the computing
centers HLRZ in J\"ulich and RUS in Stuttgart.

\clearpage
\newpage

\clearpage
\newpage

\end{document}